\documentclass[11pt,english,aps,preprint,nofootinbib,superscriptaddress]{revtex4}
\pdfoutput=1
\usepackage[T1]{fontenc}
\usepackage[latin9]{inputenc}
\usepackage{verbatim}
\usepackage{float}
\usepackage{units}
\usepackage{amsmath}
\usepackage{graphicx}

\makeatletter
\@ifundefined{textcolor}{}
{%
 \definecolor{BLACK}{gray}{0}
 \definecolor{WHITE}{gray}{1}
 \definecolor{RED}{rgb}{1,0,0}
 \definecolor{GREEN}{rgb}{0,1,0}
 \definecolor{BLUE}{rgb}{0,0,1}
 \definecolor{CYAN}{cmyk}{1,0,0,0}
 \definecolor{MAGENTA}{cmyk}{0,1,0,0}
 \definecolor{YELLOW}{cmyk}{0,0,1,0}
 }

\usepackage{amsthm}

\makeatother

\usepackage{babel}

\makeatother

\usepackage{babel}

\begin{document}
\preprint{MIT-CTP 4076}
\preprint{CERN-PH-TH-2009/175}
\title{Quantum Adiabatic Algorithms, Small Gaps, and Different Paths}

\author{Edward Farhi}
\affiliation{Center for Theoretical Physics, Massachusetts Institute
of Technology, Cambridge, MA 02139}

\author{Jeffrey Goldstone}
\affiliation{Center for Theoretical Physics, Massachusetts Institute
of Technology, Cambridge, MA 02139}

\author{David Gosset}
\affiliation{Center for Theoretical Physics, Massachusetts Institute
of Technology, Cambridge, MA 02139}

\author{\\Sam Gutmann}
\affiliation{Department of Mathematics, Northeastern University,
Boston, MA 02115}

\author{Harvey B. Meyer}
\affiliation{Center for Theoretical Physics, Massachusetts Institute
of Technology, Cambridge, MA 02139}
\affiliation{Physics Department, CERN, 1211 Geneva 23, Switzerland}

\author{Peter Shor}
\affiliation{Center for Theoretical Physics, Massachusetts Institute
of Technology, Cambridge, MA 02139}
\affiliation{Department of Mathematics, Massachusetts Institute
of Technology, Cambridge, MA 02139}

\begin{abstract}
We construct a set of instances of 3SAT which are not solved efficiently
using the simplest quantum adiabatic algorithm. These instances are
obtained by picking random clauses all consistent with two disparate
planted solutions and then penalizing one of them with a single additional
clause. We argue that by randomly modifying the beginning Hamiltonian,
one obtains (with substantial probability) an adiabatic path that
removes this difficulty. This suggests that the quantum adiabatic
algorithm should in general be run on each instance with many different
random paths leading to the problem Hamiltonian. We do not know whether
this trick will help for a random instance of 3SAT (as opposed to
an instance from the particular set we consider), especially if the
instance has an exponential number of disparate assignments that violate
few clauses. We use a continuous imaginary time Quantum Monte Carlo
algorithm in a novel way to numerically investigate the ground state
as well as the first excited state of our system. Our arguments are
supplemented by Quantum Monte Carlo data from simulations with up
to 150 spins. 
\end{abstract}
\maketitle

\section{Introduction}

Quantum adiabatic algorithms are designed for classical combinatorial
optimization problems \cite{farhi-2000}. In the simplest case, such
algorithms work by adiabatically evolving in the ground state of a
system with Hamiltonian $H(s)=(1-s)H_{B}+sH_{P}$ that is a function
of a parameter $s$ which is increased from $0$ to $1$ as a function
of time. $H_{B}$ is called the beginning Hamiltonian and $H_{P}$,
which is instance dependent, is called the problem Hamiltonian. The
minimum (for $s\in[0,1]$) eigenvalue gap between the ground state
and first excited state of $H(s)$ is related to the runtime of the
adiabatic algorithm. If $H(s)$ has an exponentially small minimum
gap then the corresponding algorithm is inefficient, whereas a minimum
gap which scales inverse polynomially corresponds to an efficient
quantum adiabatic algorithm.

Whether or not quantum adiabatic algorithms can be used to solve classically
difficult optimization problems efficiently remains to be seen. Some
numerical studies have examined the quantum adiabatic algorithm on
random sets of instances of optimization problems where these sets
are thought to be difficult for classical algorithms. These studies
have reported polynomial scaling of the minimum gap out to about $100$
bits \cite{farhi-2001,young-2008-101,hogg-2003-67}. Whether this
scaling persists at high bit number has recently been called into
question \cite{young-2009}. Meanwhile there has been no rigorous
analytical result that characterizes the performance of the quantum
adiabatic algorithms on random instances of NP-complete problems.

Over the years there have been a number of proposed examples which
were meant to demonstrate failures of the adiabatic algorithm on specific
problems. In reference \cite{vandam-2002}, van Dam et al constructed
examples intended to foil the adiabatic algorithm, but these examples
used a nonlocal cost function. Related 3SAT examples of van Dam and
Vazirani \cite{vandam-} indeed cannot be solved efficiently using
the quantum adiabatic algorithm. However it was shown in \cite{farhi-2002}
that such 3SAT instances do not pose a problem for the quantum adiabatic
algorithm if, having fixed a specific problem Hamiltonian, one randomly
chooses multiple interpolating paths between the initial and final
Hamiltonians and runs the adiabatic algorithm once for each random
path. Fisher \cite{Fisher} has constructed an interesting but specialized
example on which the quantum adiabatic algorithm is inefficient. (A
later example of Reichardt \cite{1007428} is based on this.) We do
not know if random path change helps here. (It is interesting to note
that in this case the runtime scales like $c^{\sqrt{n}}$ where $n$
is the number of bits.) The authors of reference \cite{znidaric-2006-73}
pointed out that a certain adiabatic algorithm for 3SAT is not efficient.
However, this was due to a perverse and avoidable nonlocal choice
of beginning Hamiltonian $H_{B}$ \cite{farhi-2008-6}.

\begin{figure}
\includegraphics[scale=0.55]{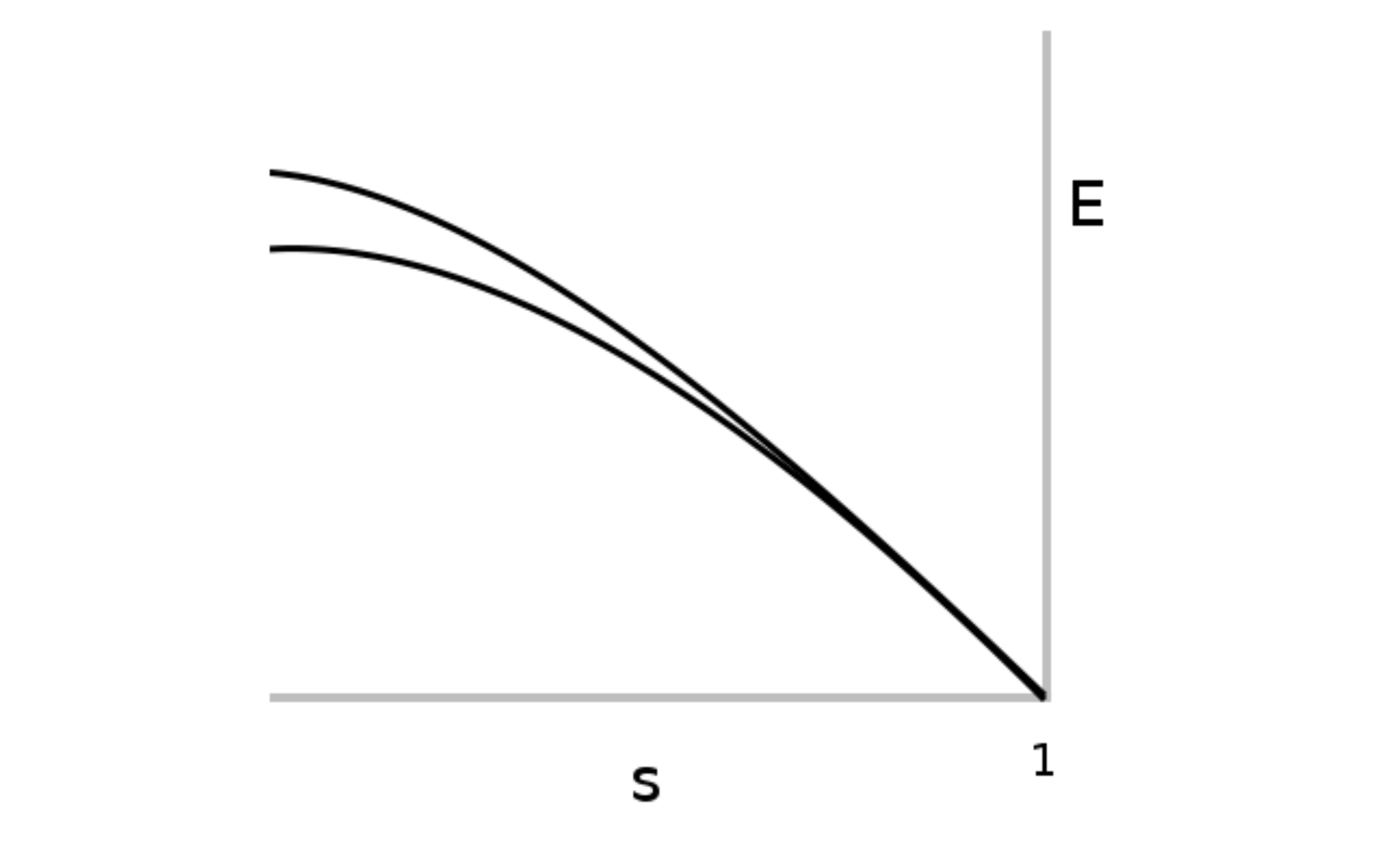}

\caption{Energy levels of the Hamiltonian $H^{\prime}(s)$ before adding the
last term $h$ to the problem Hamiltonian. The lower curve coincides
with the upper curve only at $s=1$.}

\label{Flo:before} 
\end{figure}

\begin{figure}
\includegraphics[scale=0.55]{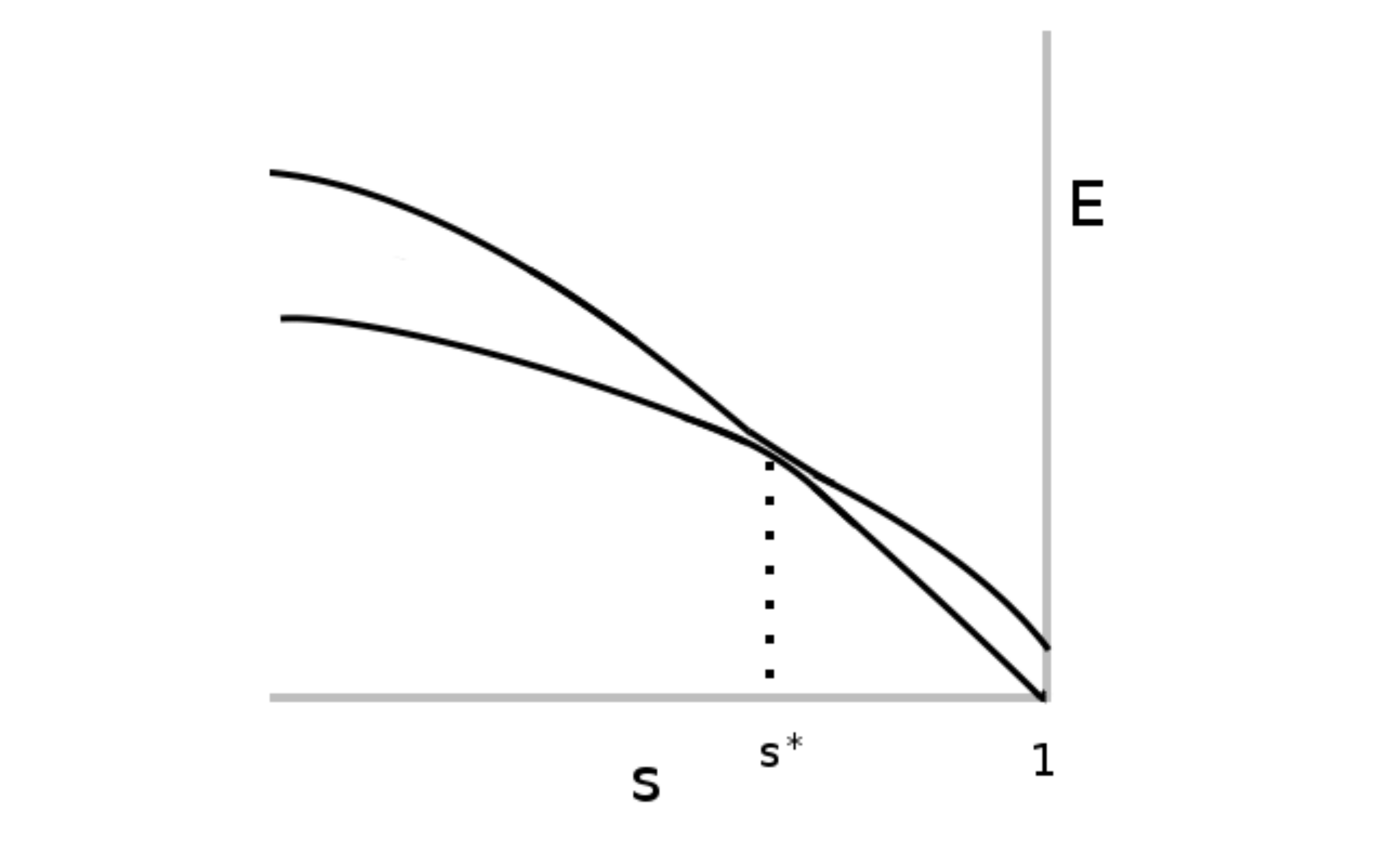}

\caption{Energy levels of the Hamiltonian $H(s)$ . There is a tiny gap at
$s^{\star}$. }

\label{Flo:after}
\end{figure}

One purpose of this paper is to discuss a different type of challenge
to adiabatic optimization which has recently come to light. (See references
\cite{amin-2009,Fan-2009,altshuler-2009} and in a different context
\cite{matsuda-2009-11}.) It seems to us that in the history of challenges
to adiabatic optimization, this may be the most serious. The takeaway
message of this work is that a small minimum gap for $H(s)$ can arise
when the Hamiltonian $H_{P}$ has features which are seen in the following
construction. First suppose that we start with a problem Hamiltonian
$H_{P}^{\prime}$ which has two degenerate ground states $|z_{1}\rangle$
and $|z_{2}\rangle$ corresponding to bit strings of length $n$ that
differ in order $n$ bits. We then expect that the two lowest eigenvalues
of $H^{\prime}(s)=(1-s)H_{B}+sH_{P}^{\prime}$ will look something
like figure \ref{Flo:before}. Note that one curve is always below
the other except at $s=1$ even though in the figure they appear to
meld because they have the same slope at $s=1$. Suppose that the
upper curve (the first excited state for $s$ near $1$) approaches
the state $|z_{2}\rangle$ and the lower curve approaches the state
$|z_{1}\rangle$ as $s\rightarrow1.$ We now form the problem Hamiltonian
$H_{P}=H_{P}^{\prime}+h$, where $h$ is a term that penalizes the
state $|z_{1}\rangle$ but not the state $|z_{2}\rangle.$ Then, for
the two lowest eigenvalues of $H(s)=(1-s)H_{B}+sH_{P}$, we expect
to have the situation pictured in figure \ref{Flo:after}, where there
is a small gap near $s=1$.

In this paper we discuss a simple way of generating random instances
of 3SAT where the corresponding adiabatic Hamiltonian $H(s)$ has
the difficulty discussed above. We argue in section \ref{sec:Fixing-the-Problem}
that this problem can be overcome (for the set of instances we consider)
by randomizing the choice of beginning Hamiltonian.

We present a continuous imaginary time Quantum Monte Carlo algorithm,
which is a modification of the Heat Bath algorithm of Krzakala et
al \cite{krzakala-2008-78}. ({}``Quantum Monte Carlo'' is a completely
classical numerical technique for finding properties of quantum systems
and is not a quantum algorithm.) We use this numerical technique in
a novel manner which allows us to investigate both the ground state
and the first excited state of our Hamiltonian and thereby detect
the presence or absence of a small gap between them. This differs
from the standard application of the Quantum Monte Carlo method in
that we are able to obtain information about the first excited state
using a simple procedure which we have validated at low bit number
by comparing our results to exact numerical diagonalization. These
Quantum Monte Carlo simulations support our claim that the problem
we describe can be overcome using random path change.

\section{Problematic Instances\label{sec:Problematic-Instances}}

We now describe the method we use to generate $n$ bit random instances
of 3SAT that lead to quantum adiabatic Hamiltonians with small minimum
gaps. To do this we first generate an instance with exactly two satisfying
assignments given by the bit strings $111...1$ and $000...0$ . Each
clause $c$ of the 3SAT instance specifies a subset of 3 bits $i_{1}(c),i_{2}(c),i_{3}(c)\in\{1,..,n\}$
and a particular assignment $w_{1}(c)w_{2}(c)w_{3}(c)$ to those three
bits which is disallowed. In order to only generate instances which
are consistent with the bit strings $111...1$ and $000...0$, we
only use clauses for which \[
w_{1}(c)w_{2}(c)w_{3}(c)\in\{100,010,001,110,101,011\}\,.\]
 We add such clauses one at a time uniformly at random and stop as
soon as these two bit strings are the only bit strings which satisfy
all of the clauses that have been added. (In practice to check whether
or not this is the case we use a classical 3SAT solver.) We write
$m$ for the total number of clauses in the instance. We note that
the number of clauses $m$ obtained using this procedure scales like
$n\log n$. We need this many clauses in order to ensure that each
bit is involved in some clause. On the other hand, when the number
of clauses is $5n\log n$ the probability of additional satisfying
assignments goes to zero as $n\rightarrow\infty$.

We now consider the problem Hamiltonian $H_{P}^{\prime}$ corresponding
to this instance, which we define to be\[
H_{P}^{\prime}=\sum_{c=1}^{m}\left(\frac{1+(-1)^{w_{1}(c)}\sigma_{z}^{i_{1}(c)}}{2}\right)\left(\frac{1+(-1)^{w_{2}(c)}\sigma_{z}^{i_{2}(c)}}{2}\right)\left(\frac{1+(-1)^{w_{3}(c)}\sigma_{z}^{i_{3}(c)}}{2}\right)\,.\]

Each term in this sum is $1$ if $i_{1}(c)i_{2}(c)i_{3}(c)$ violates
the clause and $0$ otherwise. For the beginning Hamiltonian we choose
\begin{equation}
H_{B}=\sum_{i=1}^{n}\left(\frac{1-\sigma_{x}^{i}}{2}\right)\,.\label{eq:Hb}\end{equation}
 The two lowest eigenvalues of the Hamiltonian $H^{\prime}(s)=(1-s)H_{B}+sH_{P}^{\prime}$
will then both approach $0$ as $s\rightarrow1$, as in figure \ref{Flo:before}.
The ground state for values of $s$ which are sufficiently close to
1 will approach either $|000...0\rangle$ or $|111...1\rangle$ as
$s\rightarrow1$. Suppose for values of $s$ close to $1$ the state
that approaches $|000...0\rangle$ has lowest energy. Then we add
an extra term $h_{0}$ which acts on bits $1,2$ and $3$ and which
penalizes this state but not $|111...1\rangle$\[
h_{0}=\frac{1}{2}\left(\frac{1+\sigma_{z}^{1}}{2}\right)\left(\frac{1+\sigma_{z}^{2}}{2}\right)\left(\frac{1+\sigma_{z}^{3}}{2}\right)\]
 and pick \[
H_{P}=H_{P}^{\prime}+h_{0}\,.\]
 (Note that this extra term has a multiplicative factor of $\frac{1}{2}$,
to avoid any degeneracy of the first excited state at $s=1$.) In
the case where the lowest energy state near $s=1$ approaches $|111...1\rangle$
we instead add\[
h_{1}=\frac{1}{2}\left(\frac{1-\sigma_{z}^{1}}{2}\right)\left(\frac{1-\sigma_{z}^{2}}{2}\right)\left(\frac{1-\sigma_{z}^{3}}{2}\right)\,.\]
 The Hamiltonian $H(s)=(1-s)H_{B}+sH_{P}$ is then expected to have
a small gap near $s=1$ as depicted in figure \ref{Flo:after}. We
expect the location $s^{\star}$ of the minimum gap to approach $s=1$
as $n\rightarrow\infty$.

\subsection*{Location of the Minimum Gap}

In order to determine the dependence of the location $s^{\star}$
of the avoided crossing on the number of spins $n$ and the number
of clauses $m$, we consider the perturbative corrections to the energies
of the states $|000...0\rangle$ and $|111...1\rangle$ around $s=1$.
We will show that the low order terms in the perturbation series reliably
predict a crossing at%
\footnote{The notation $f(n)=\Theta(g(n))$ means that, for $n$ sufficiently
large, $bg(n)\leq f(n)\leq cg(n)$ for some constants $b$ and $c$. %
} $s^{\star}=1-\Theta(\frac{1}{n^{\nicefrac{1}{4}}}\left(\frac{m}{n}\right)^{\frac{3}{4}})$
.

With the Hamiltonian constructed in the previous section, one of these
states has energy $0$ at $s=1$ (call this the lower state $|z_{L}\rangle$)
and the other state has energy $\frac{1}{2}$ at $s=1$ (we call this
the upper state $|z_{U}\rangle$). We write our Hamiltonian as\begin{equation}
H(s)=(1-s)\frac{n}{2}+s\left[-\left(\frac{1-s}{s}\right)\sum_{i=1}^{n}\frac{\sigma_{x}^{i}}{2}+H_{P}\right]\label{eq:Ham}\end{equation}
 and we then consider the term $-\left(\frac{1-s}{s}\right)\sum_{i=1}^{n}\frac{\sigma_{x}^{i}}{2}$
as a perturbation to $H_{P}$ expanding around $s=1$.

\begin{figure}
\includegraphics[scale=0.55]{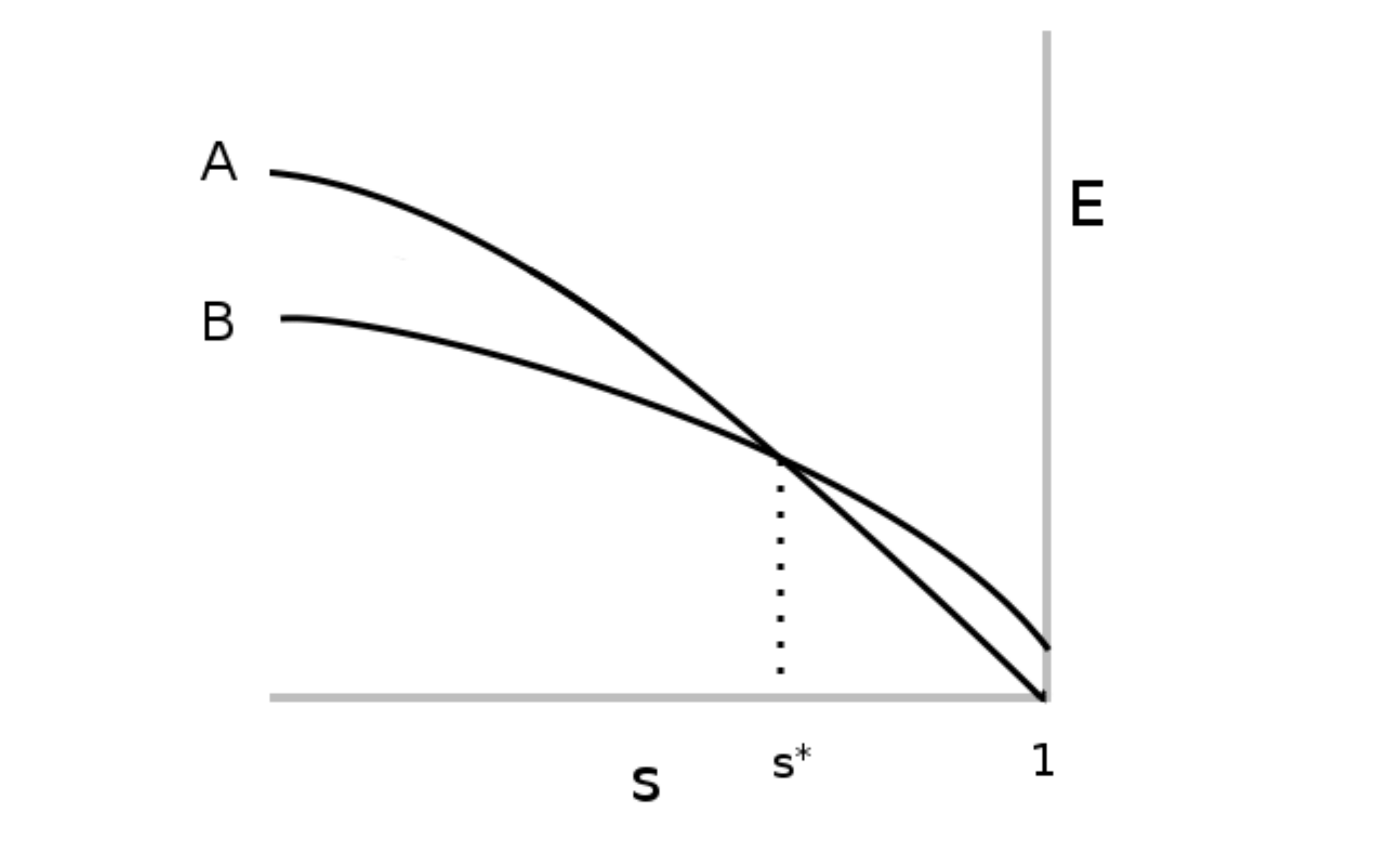}

\caption{A true energy level crossing can arise from two disconnected sectors. }

\label{Flo:realcross} 
\end{figure}

To understand why we trust perturbation theory to predict the location
of the near crossing, consider a system which is composed of two disconnected
sectors, $A$ and $B$, so the corresponding Hamiltonian is of the
form\[
\left(\begin{array}{cc}
H_{A}(s) & 0\\
0 & H_{B}(s)\end{array}\right)\,.\]
 In this situation the generic rule that levels do not cross does
not apply and we can easily imagine that the two lowest levels look
like what we show in figure \ref{Flo:realcross}, where the levels
actually cross at $s^{\star}$.

Imagine that low order perturbation theory around $s=1$ can be used
to get good approximations to the ground state energies of $H_{A}(s)$
and $H_{B}(s)$ for $s$ near $s^{\star}$, even for $s$ somewhat
to the left of $s^{\star}.$ Then it is possible to accurately predict
$s^{\star}$. Our situation is very close to this. We can think of
$A$ as consisting of the states close to $z_{L}$ in Hamming weight,
and $B$ as states close to $z_{U}$ in Hamming weight. Similarly,
we view $H_{A}$ and $H_{B}$ as the restrictions of $H$ to these
sectors. Note that it takes $n$ powers of the perturbation to connect
$|000...0\rangle$ and $|111...1\rangle$ and this is why we view
$A$ and $B$ as essentially disconnected.

Although figure \ref{Flo:realcross} looks like figure \ref{Flo:after},
it is figure \ref{Flo:after} that depicts the actual situation, where
the two levels avoid crossing. This true near cross means that the
perturbation series in the actual theory will diverge very close to
$s^{\star}$. However this divergence will only be seen at high order,
in fact at an order which is proportional to $n$. The low order terms
of the perturbation series in figure \ref{Flo:realcross} are the
same as the low order terms of the perturbation series in figure \ref{Flo:after},
so we can trust low order perturbation theory to locate $s^{\star}$.

(We argue below that as a function of the number of bits, $n$, $s^{\star}$
goes to $1$ as $n$ goes to infinity. This implies that the radius
of convergence of the perturbation theory for the full $H(s)$, expanded
about $s=1$, goes to $0$ as $n$ goes to infinity. This fact has
no bearing on our argument that low order perturbation theory can
be used to accurately predict $s^{\star}.)$

For small values of the parameter $\frac{1-s}{s}$, the energies of
the two states under consideration can be expanded as\begin{eqnarray}
E_{L}(s) & = & (1-s)\frac{n}{2}+s\left[0+\left(\frac{1-s}{s}\right)^{2}e_{L}^{(2)}+\left(\frac{1-s}{s}\right)^{4}e_{L}^{(4)}+...\right]\label{eq:EL(s)}\\
E_{U}(s) & = & (1-s)\frac{n}{2}+s\left[\frac{1}{2}+\left(\frac{1-s}{s}\right)^{2}e_{U}^{(2)}+\left(\frac{1-s}{s}\right)^{4}e_{U}^{(4)}+...\right]\,.\label{eq:EU(s)}\end{eqnarray}
 It is easy to see that each expansion (inside the square brackets)
only contains even powers. Note that $e_{L}^{(2)}$ is guaranteed
to be negative as is always the case for 2nd order perturbation theory
of the ground state. In addition, since we added a term to penalize
the state which had smaller energy near $s=1$ before adding the clause,
we expect that $e_{U}^{(2)}<e_{L}^{(2)}$. We are interested in the
behaviour of the difference $E_{L}(s)-E_{U}(s)$ for randomly generated
instances (as generated using the prescription of the previous section)
as a function of the number of spins $n$ in the limit $n\rightarrow\infty$.
This requires us to further investigate the behaviour of each coefficient
$e_{L}^{(k)}$ and $e_{U}^{(k)}$ as a function of $n$ and $m$.

In fact to locate $s^{\star}$ we need only to go to second order
in perturbation theory. From equation \ref{eq:Ham} we view $H_{P}$
as the unperturbed Hamiltonian, and $-\frac{1}{2}\sum_{i=1}^{n}\sigma_{x}^{i}$
as the perturbation, with $\frac{1-s}{s}$ as the expansion parameter.
Now $\sigma_{x}^{i}|z\rangle=|z\oplus\hat{e}_{i}\rangle$ so at second
order we get\begin{eqnarray*}
e_{L}^{(2)} & = & \frac{1}{4}\sum_{i=1}^{n}\frac{|\langle z_{L}\oplus\hat{e}_{i}|\sigma_{x}^{i}|z_{L}\rangle|^{2}}{\langle z_{L}|H_{P}|z_{L}\rangle-\langle z_{L}\oplus\hat{e}_{i}|H_{P}|z_{L}\oplus\hat{e}_{i}\rangle}\\
 & = & -\frac{1}{4}\sum_{i=1}^{n}\frac{1}{\langle z_{L}\oplus\hat{e}_{i}|H_{P}|z_{L}\oplus\hat{e}_{i}\rangle}\,.\end{eqnarray*}
 Similarly, \[
e_{U}^{(2)}=\frac{1}{4}\sum_{i=1}^{n}\frac{1}{\frac{1}{2}-\langle z_{U}\oplus\hat{e}_{i}|H_{P}|z_{U}\oplus\hat{e}_{i}\rangle}\;\;\;\;\;\;\;\qquad\]
 where the $\frac{1}{2}$ in the denominator is $\langle z_{U}|H_{P}|z_{U}\rangle.$

The expected energy penalty incurred when flipping a bit of either
$z_{U}$ or $z_{L}$ is of order $\frac{m}{n}$ since each bit is
typically involved in $\Theta(\frac{m}{n})$ clauses. So the coefficients
$e_{L}^{(2)}$ and $e_{U}^{(2)}$ are of order $n\left(\frac{n}{m}\right)$
since the energy denominators involved are $\Theta(\frac{m}{n})$.
We now show that their difference is of order $\sqrt{n}\left(\frac{n}{m}\right)^{\frac{3}{2}}$.
Write \[
e_{U}^{(2)}-e_{L}^{(2)}=\sum_{i=1}^{n}d_{i}\]
 where for each $i$, \begin{equation}
d_{i}=\frac{1}{4}\bigg(\frac{1}{\frac{1}{2}-\langle z_{U}\oplus\hat{e}_{i}|H_{P}|z_{U}\oplus\hat{e}_{i}\rangle}+\frac{1}{\langle z_{L}\oplus\hat{e}_{i}|H_{P}|z_{L}\oplus\hat{e}_{i}\rangle}\bigg)\,.\label{eq:d_i}\end{equation}

Recall that $H_{P}=H_{P}^{\prime}+h$ where $h$ is the penalty term
from the final clause which acts only on the first 3 bits. Therefore,
for $i=4,5,...,n$\[
d_{i}=\frac{1}{4}\bigg(-\frac{1}{\langle z_{U}\oplus\hat{e}_{i}|H_{P}^{\prime}|z_{U}\oplus\hat{e}_{i}\rangle}+\frac{1}{\langle z_{L}\oplus\hat{e}_{i}|H_{P}^{\prime}|z_{L}\oplus\hat{e}_{i}\rangle}\bigg)\,.\]
 Our procedure for generating instances is symmetric between the strings
$000...0$ and $111...1$ so averaging over instances it is clear
that the mean of $d_{i}$ for $i=4,5...,n$ is $0$. Thus we expect
$\sum_{i=4}^{n}d_{i}$ to be (approximately) Gaussian with mean $0$
and standard deviation proportional to $\sqrt{n}\sigma(d)$, where
$\sigma(d)$ is the standard deviation of each $d_{i}$ for $i\in\{4,5,...,n\}$.
To compute $\sigma(d)$ we note that \[
d_{i}=\frac{1}{4}\bigg(\frac{\langle z_{U}\oplus\hat{e}_{i}|H_{P}^{\prime}|z_{U}\oplus\hat{e}_{i}\rangle-\langle z_{L}\oplus\hat{e}_{i}|H_{P}^{\prime}|z_{L}\oplus\hat{e}_{i}\rangle}{\langle z_{L}\oplus\hat{e}_{i}|H_{P}^{\prime}|z_{L}\oplus\hat{e}_{i}\rangle\langle z_{U}\oplus\hat{e}_{i}|H_{P}^{\prime}|z_{U}\oplus\hat{e}_{i}\rangle}\bigg)\,.\]
 Again using the symmetry between all zeros and all ones, we conclude
that the numerator is of order $\sqrt{\frac{m}{n}}$ and the denominator
is of order $\left(\frac{m}{n}\right)^{2}$. Hence we expect $\sigma(d)$
to be $\Theta\big(\left(\frac{n}{m}\right)^{\frac{3}{2}}\big)$. So
$e_{U}^{(2)}-e_{L}^{(2)}$ is of order $\sqrt{n}\left(\frac{n}{m}\right)^{\frac{3}{2}}.$
We will now locate $s^{\star}$ using second order perturbation theory
and afterwards argue that higher orders do not change the result.
Returning to equations \ref{eq:EL(s)} and \ref{eq:EU(s)}, equating
the two energies at second order we have\[
0=s^{\star}\left[\frac{1}{2}+\left(\frac{1-s^{\star}}{s^{\star}}\right)^{2}\left(e_{U}^{(2)}-e_{L}^{(2)}\right)\right]\]
 so \begin{equation}
s^{\star}=1-\Theta(\frac{1}{n^{\nicefrac{1}{4}}}\left(\frac{m}{n}\right)^{\frac{3}{4}})\,.\label{eq:s_star}\end{equation}

The 4th order correction to the energy of the lower state is given
by\[
e_{L}^{(4)}=\langle z_{L}|V\big(\frac{\phi_{L}}{H_{P}}\big)^{2}V|z_{L}\rangle\langle z_{L}|V\frac{\phi_{L}}{H_{P}}V|z_{L}\rangle-\langle z_{L}|V\frac{\phi_{L}}{H_{P}}V\frac{\phi_{L}}{H_{P}}V\frac{\phi_{L}}{H_{P}}V|z_{L}\rangle\]
 where \[
V=-\frac{1}{2}\sum_{i=1}^{n}\sigma_{x}^{i}\]
 and $\phi_{L}=1-|z_{L}\rangle\langle z_{L}|$. Writing $\mathcal{H}_{P}(z)=\langle z|H_{P}|z\rangle$,
this can be expressed as\begin{eqnarray*}
e_{L}^{(4)} & = & \frac{1}{16}\sum_{i=1}^{n}\sum_{j=1}^{n}\frac{1}{\left(\mathcal{H}_{P}(z_{L}\oplus\hat{e}_{i})\right)^{2}\mathcal{H}_{P}(z_{L}\oplus\hat{e}_{j})}\\
 &  & -\frac{1}{16}\sum_{i\neq j}\frac{1}{\mathcal{H}_{P}(z_{L}\oplus\hat{e}_{i})\mathcal{H}_{P}(z_{L}\oplus\hat{e}_{j})\mathcal{H}_{P}(z_{L}\oplus\hat{e}_{i}\oplus\hat{e}_{j})}\\
 &  & -\frac{1}{16}\sum_{i\neq j}\frac{1}{\left(\mathcal{H}_{P}(z_{L}\oplus\hat{e}_{i})\right)^{2}\mathcal{H}_{P}(z_{L}\oplus\hat{e}_{i}\oplus\hat{e}_{j})}\\
 & = & \sum_{i=1}^{n}\frac{1}{16}\frac{1}{\left(\mathcal{H}_{P}(z_{L}\oplus\hat{e}_{i})\right)^{3}}+\sum_{i\neq j}\frac{1}{16}\frac{\mathcal{H}_{P}(z_{L}\oplus\hat{e}_{i}\oplus\hat{e}_{j})-\mathcal{H}_{P}(z_{L}\oplus\hat{e}_{i})-\mathcal{H}_{P}(z_{L}\oplus\hat{e}_{j})}{\left(\mathcal{H}_{P}(z_{L}\oplus\hat{e}_{i})\right)^{2}\mathcal{H}_{P}(z_{L}\oplus\hat{e}_{j})\left(\mathcal{H}_{P}(z_{L}\oplus\hat{e}_{i}\oplus\hat{e}_{j})\right)}\,.\end{eqnarray*}
 Now consider the terms in this expression corresponding to indices
$i,j$ for which $i\neq j$ and the bits $i$ and $j$ do not appear
in a clause together. Under these conditions we have \[
\mathcal{H}_{P}(z_{L}\oplus\hat{e}_{i}\oplus\hat{e}_{j})=\mathcal{H}_{P}(z_{L}\oplus\hat{e}_{i})+\mathcal{H}_{P}(z_{L}\oplus\hat{e}_{j})\,.\]
 So we can write \begin{eqnarray}
e_{L}^{(4)} & = & \sum_{i=1}^{n}\frac{1}{16}\frac{1}{\left(\mathcal{H}_{P}(z_{L}\oplus\hat{e}_{i})\right)^{3}}\label{eq:E4L}\\
 &  & +\sum_{i\neq j\text{ clausemates}}\frac{1}{16}\frac{\mathcal{H}_{P}(z_{L}\oplus\hat{e}_{i}\oplus\hat{e}_{j})-\mathcal{H}_{P}(z_{L}\oplus\hat{e}_{i})-\mathcal{H}_{P}(z_{L}\oplus\hat{e}_{j})}{\left(\mathcal{H}_{P}(z_{L}\oplus\hat{e}_{i})\right)^{2}\mathcal{H}_{P}(z_{L}\oplus\hat{e}_{j})\left(\mathcal{H}_{P}(z_{L}\oplus\hat{e}_{i}\oplus\hat{e}_{j})\right)}\,.\nonumber \end{eqnarray}
 Here the subscript {}``clausemates'' indicates that we sum only
over pairs of indices which appear together in at least one clause
of the 3SAT instance corresponding to $H_{P}$. For $e_{U}^{(4)}$,
since the unperturbed energy is $\frac{1}{2}$ we obtain\begin{eqnarray}
e_{U}^{(4)} & = & \sum_{i=1}^{n}\frac{1}{16}\frac{1}{\left(\mathcal{H}_{P}(z_{U}\oplus\hat{e}_{i})-\frac{1}{2}\right)^{3}}\label{eq:E4U}\\
 &  & +\sum_{i\neq j\text{ clausemates}}\frac{1}{16}\frac{\left(\mathcal{H}_{P}(z_{U}\oplus\hat{e}_{i}\oplus\hat{e}_{j})-\frac{1}{2}\right)-\left(\mathcal{H}_{P}(z_{U}\oplus\hat{e}_{i})-\frac{1}{2}\right)-\left(\mathcal{H}_{P}(z_{U}\oplus\hat{e}_{j})-\frac{1}{2}\right)}{\left(\mathcal{H}_{P}(z_{U}\oplus\hat{e}_{i})-\frac{1}{2}\right)^{2}\left(\mathcal{H}_{P}(z_{U}\oplus\hat{e}_{j})-\frac{1}{2}\right)\left(\mathcal{H}_{P}(z_{U}\oplus\hat{e}_{i}\oplus\hat{e}_{j})-\frac{1}{2}\right)}\,.\nonumber \end{eqnarray}

Let's look at the first sum in each of equations \ref{eq:E4L} and
\ref{eq:E4U} where $i$ goes from $1$ to $n$. Each term as $i$
goes from $4$ to $n$ is of order $\left(\frac{n}{m}\right)^{3}$
and so the difference of the two sums is $\Theta(\sqrt{n}\left(\frac{n}{m}\right)^{3})$
. The second sums (those which are restricted to clausemates) contain
of order $m$ terms. Each denominator is of order $\left(\frac{m}{n}\right)^{4}$
and the numerators are $\Theta(1)$ since the only contribution to
the numerator is from clauses in $H_{P}$ which involve bits $i$
and $j$ together. Separating out the terms where $i$ and/or $j$
are $1,2$ or $3$, we conclude that the contribution to $e_{U}^{(4)}-e_{L}^{(4)}$
from the clausemate terms is $\Theta(\sqrt{n}\left(\frac{n}{m}\right)^{\frac{7}{2}})$.
For our instance generation $m$ grows like $n\log n$ so this clausemate
contribution is asymptotically dominated by the first term which scales
like $\Theta(\sqrt{n}\left(\frac{n}{m}\right)^{3})$. So the fourth
order contribution to the difference of energies $E_{U}(s)-E_{L}(s)$
is $\Theta(s\left[\sqrt{n}\left(\frac{n}{m}\right)^{3}\left(\frac{1-s}{s}\right)^{4}\right])$.
At $s^{\star}$ which we determined at second order to be $1-\Theta(\frac{1}{n^{\nicefrac{1}{4}}}\left(\frac{m}{n}\right)^{\frac{3}{4}})$,
the fourth order contribution to the energy difference is $\Theta\left(\frac{1}{\sqrt{n}}\right)$.
The fourth order corrections can therefore be neglected in determining
the location of $s^{\star}.$ Sixth order and higher contributions
to the difference are even smaller.

\section{Fixing the Problem by Path Change\label{sec:Fixing-the-Problem}}

\begin{figure}
\begin{centering}
\includegraphics[scale=0.55]{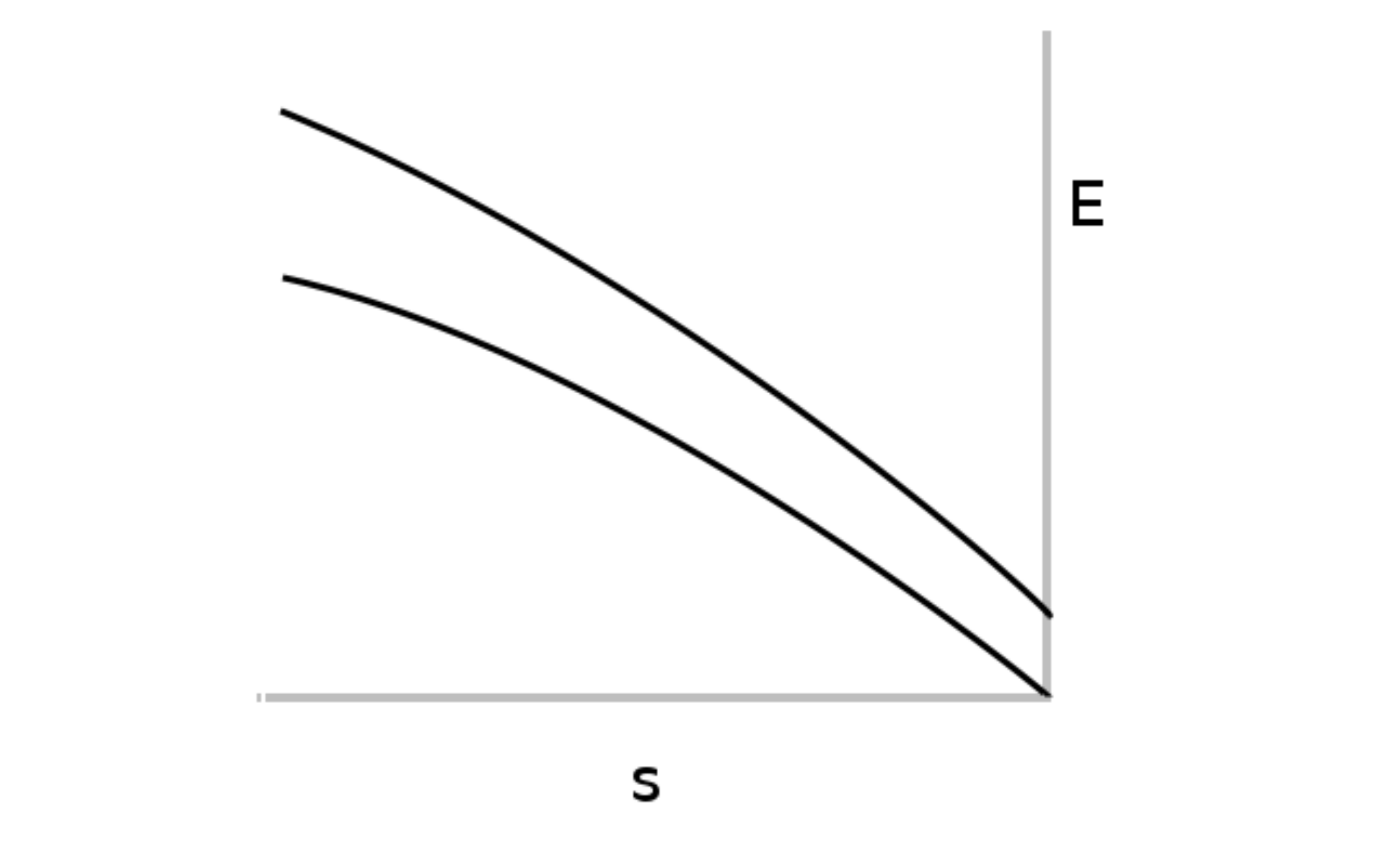} 
\par\end{centering}

\caption{Energy levels with no avoided crossing near $s=1$. Here the second
derivative of the upper curve B is greater than the second derivative
of the lower curve A.}

\label{Flo:fixed}
\end{figure}

In our instance generation we manufactured a small gap by penalizing
the planted assignment corresponding to the energy eigenstate with
the smallest energy near $s=1$. Since the slopes of the two curves
in figure \ref{Flo:before} are the same at $s=1$, the second derivatives
determine which eigenvalue is smaller near $s=1$. After penalization
we have $e_{U}^{(2)}<e_{L}^{(2)}$ which is consistent with the near
crossing in figure \ref{Flo:after}. Suppose instead that we penalize
the assignment corresponding to larger energy in figure \ref{Flo:before}.
Then we expect the situation depicted in figure \ref{Flo:fixed} where
no level crossing is induced.

We imagine that the instances that we manufacture with a small gap
as in figure \ref{Flo:after} are a model for what might be encountered
in running the quantum adiabatic algorithm on some instance a quantum
computer is actually trying to solve. There is a strategy for overcoming
this problem. The idea is to produce figure \ref{Flo:fixed}, with
reasonable probability, by randomly modifying the adiabatic path which
ends at $H_{P}$, of course making no use of the properties of the
particular instance. For this purpose one could use any random ensemble
of paths $H(s)$ such that $H(1)=H_{P}$ and the ground state of $H(0)$
is simple to prepare. However in this paper we only consider randomly
changing the beginning Hamiltonian. We have made this choice so that
we are able to use our Quantum Monte Carlo method to numerically verify
our arguments. Like other Quantum Monte Carlo methods, the method
we use does not work when the Hamiltonian has nonzero off-diagonal
matrix elements with positive sign.

Starting with an instance where $H(s)=(1-s)H_{B}+sH_{P}$ has a tiny
gap due to the problem discussed above, we now consider a different
adiabatic path $\tilde{H}(s)=(1-s)\tilde{H}_{B}+sH_{P}$ obtained
by keeping the same problem Hamiltonian $H_{P}$ but choosing a different
beginning Hamiltonian $\tilde{H}_{B}$ in a random fashion which we
will prescribe below. We argue that the small gap near $s=1$ is then
removed with substantial probability, so that by repeating this procedure
a constant number of times it is possible to find an adiabatic path
without a small gap near $s=1$. %

The way that we choose a random Hamiltonian $\tilde{H}_{B}$ is to
first draw $n$ random variables $c_{i}$ for $i=1,2,...,n$ , where
each $c_{i}$ is chosen to be $\frac{1}{2}$ or $\frac{3}{2}$ with equal probability.
We then take\begin{equation}
\tilde{H}_{B}=\sum_{i=1}^{n}\frac{c_{i}(1-\sigma_{x}^{i})}{2}\,.\label{eq:tildeHb}\end{equation}
 We write $\tilde{e}_{U}^{(2)}$ and $\tilde{e}_{L}^{(2)}$ for the
analogous quantities to $e_{L}^{(2)}$ and $e_{U}^{(2)}$ for the
new Hamiltonian $\tilde{H}(s)=(1-s)\tilde{H}_{B}+sH_{P}$. The point
is that by randomizing $\tilde{H}_{B}$ in the way we prescribe, there
is a substantial probability that one will obtain $\tilde{e}_{U}^{(2)}-\tilde{e}_{L}^{(2)}>0$
, and in that case one expects no avoided crossing near $s=1.$ Write
\[
\tilde{e}_{U}^{(2)}-\tilde{e}_{L}^{(2)}=\sum_{i=1}^{n}c_{i}^{2}d_{i}\]
 where the $\{d_{i}\}$ are fixed by the instance (and are defined
in equation \ref{eq:d_i}). Since we have fixed the problem Hamiltonian
$H_{P}$, the only random variables appearing in the above equation
are the $c_{i}$. We have $\overline{c_{i}^{2}}=\frac{5}{4}$ so the mean
value of $\tilde{e}_{U}^{(2)}-\tilde{e}_{L}^{(2)}$ is then \[
\overline{\tilde{e}_{U}^{(2)}-\tilde{e}_{L}^{(2)}}=\frac{5}{4}(e_{U}^{(2)}-e_{L}^{(2)})<0\,.\]
 But more importantly \[
\overline{\tilde{e}_{U}^{(2)}-\tilde{e}_{L}^{(2)}}=\Theta(\sqrt{n}\left(\frac{n}{m}\right)^{\frac{3}{2}})\,.\]
 The variance of this difference is \begin{eqnarray*}
\text{Var}\bigg(\tilde{e}_{U}^{(2)}-\tilde{e}_{L}^{(2)}\bigg) & = & \sum_{i=1}^{n}d_{i}^{2}\text{Var}(c_{i}^{2})\\
 & = & \sum_{i=1}^{n}d_{i}^{2}\cdot1\,,\end{eqnarray*}
 which is $\Theta(n\left(\frac{n}{m}\right)^{3})$. For a fixed instance
with a corresponding fixed set $\{d_{i}\}$ the random variable $\sum_{i}c_{i}^{2}d_{i}$
is approximately Gaussian and from its mean and variance we see that
the probability that $\tilde{e}_{U}^{(2)}-\tilde{e}_{L}^{(2)}$ is
positive and in fact greater than $a\sqrt{n}\left(\frac{n}{m}\right)^{\frac{3}{2}}$,
for $a>0$, is bounded away from $0$ independent of $n$. This means
that there is a good chance that randomizing $H_{B}$ turns the situation
depicted in figure \ref{Flo:after} into the situation depicted in
figure \ref{Flo:fixed}.

In the case of two planted satisfying assignments with one penalized
to produce a small gap when the beginning Hamiltonian is $H_{B}$
of equation \ref{eq:Hb}, we have shown that a random choice for the
beginning Hamiltonian $\tilde{H}_{B}$ of equation \ref{eq:tildeHb}
can with substantial probability remove the small gap. This gives
further weight to the idea that when running the quantum adiabatic
algorithm on a single instance of some optimization problem, the programmer
should run the quantum adiabatic algorithm repeatedly with different
paths ending at $H_{P}$ \cite{farhi-2002}.

\subsection*{What if there are $k>2$ satisfying assignments and all but one are
penalized?}

We have considered instances of 3SAT for which the corresponding problem
Hamiltonian has a unique ground state and a nondegenerate first excited
state that is far from the ground state in Hamming weight. In this
case we have shown that the path change strategy succeeds in removing
a tiny gap. What happens when we have $k$ mutually disparate solutions
and we penalize all but one of them? We show (under some assumptions) that for large $n$,
path change will succeed after a number of tries polynomial in $k$.
We can therefore hope for success if there are $k$ disparate assignments
which violate few clauses and $k$ scales polynomially with $n$.
On the other hand when $k$ is superpolynomial in $n$ we have no
reason to be optimistic about the performance of the quantum adiabatic
algorithm.

The energies $E_{r}(s)$ for $r=0,1,..,k-1$ of the $k$
states under consideration can be expanded as in equations \ref{eq:EL(s)}
and \ref{eq:EU(s)}. Write $e_{r}^{(2)}$ for the second order corrections
in these expansions. If we add a small number of clauses to this instance which penalize
all the solutions except for a randomly chosen one which we call $z_{0}$ (forming a new problem Hamiltonian
$\hat{H}_{p}$) then the differences $e_{r}^{(2)}-e_{0}^{(2)}$ will
not change substantially. So we are interested in the differences \[
e_{r}^{(2)}-e_{0}^{(2)}=\sum_{i=1}^{n} d_{r i},\]
where \begin{equation}
d_{r i}=\frac{1}{4}\left(-\frac{1}{\langle z_{r}\oplus\hat{e}_{i}|H_{p}|z_{r}\oplus\hat{e}_{i}\rangle}+\frac{1}{\langle z_{0}\oplus\hat{e}_{i}|H_{P}|z_{0}\oplus\hat{e}_{i}\rangle}\right),\label{eq:energy_denom_diff}\end{equation}
and where $z_0,z_1,z_2,...,z_{k-1}$  are the $k$ ground states of $H_p$.
The adiabatic algorithm applied to this instance with problem Hamiltonian $\hat{H}_{p}$ will succeed if all
of the above differences are positive, for $r=1,...,k-1.$ There is a $\frac{1}{k}$ chance of this occurring. If they are not all positive, then we have encountered an instance which we expect to have a small gap. In this case, we randomize the beginning Hamiltonian as in
the previous subsection. This produces new values $\tilde{e}_{r}^{(2)}$
for the second order energy corrections and the relevant differences
are given by \begin{equation}
\tilde{e}_{r}^{(2)}-\tilde{e}_{0}^{(2)}=\sum_{i=1}^{n}c_{i}^{2}d_{r i},\label{eq:tilde_diff}\end{equation}
where the random variables $c_{i}$ are $\frac{1}{2}$ or $\frac{3}{2}$
with equal probability (and the $d_{r i}$ are fixed). With this fixed problem Hamiltonian, how many times do we have to randomize the beginning Hamiltonian (by drawing a random set $\{c_i\}$) to produce algorithmic success? For large
$n$, the random variables $\tilde{e}_{r}^{(2)}-\tilde{e}_{0}^{(2)}$
are approximately jointly Gaussian with expectation and variance given by\begin{eqnarray*}
\overline{\tilde{e}_{r}^{(2)}-\tilde{e}_{0}^{(2)}} & = & \frac{5}{4}\sum_{i=1}^{n}d_{r i}\\
\text{Var}\left(\tilde{e}_{r}^{(2)}-\tilde{e}_{0}^{(2)}\right) & = & \sum_{i=1}^{n}\left(d_{r i}\right)^{2}.\end{eqnarray*}
 The correlation (the covariance divided by the product of the standard
deviations) between any two of these random variables is given by\begin{equation}
\text{Corr }\left(\tilde{e}_{q}^{(2)}-\tilde{e}_{0}^{(2)},\tilde{e}_{r}^{(2)}-\tilde{e}_{0}^{(2)}\right)=\frac{\sum_{i}d_{qi}d_{r i}}{\left[\sum_{j}\left(d_{qj}\right)^{2}\sum_{j}\left(d_{rj}\right)^{2}\right]^{\frac{1}{2}}}.\label{eq:correlation}\end{equation}
 We now argue that the above correlation is typically equal to $\frac{1}{2}$
when $n$ is large. To see this, we can expand the above expression
using equation \ref{eq:energy_denom_diff}. We imagine that the instances have been generated so that the law of large numbers applies and that in the limit of large n the quantity \begin{equation*}
\frac{1}{n}\sum_{i=1}^{n}\left(\frac{1}{\langle z_{r}\oplus\hat{e}_{i}|H_{p}|z_{r}\oplus\hat{e}_{i}\rangle}\cdot\frac{1}{\langle z_{r}\oplus\hat{e}_{i}|H_{p}|z_{r}\oplus\hat{e}_{i}\rangle}\right) \\
\end{equation*}
approaches an $r$ independent value and the quantity 
\begin{equation*}
 \frac{1}{n}\sum_{i=1}^{n}\left(\frac{1}{\langle z_{q}\oplus\hat{e}_{i}|H_{p}|z_{q}\oplus\hat{e}_{i}\rangle}\cdot\frac{1}{\langle z_{r}\oplus\hat{e}_{i}|H_{p}|z_{r}\oplus\hat{e}_{i}\rangle}\right) \\
\end{equation*}
approaches a value independent of $r$ and $q$ for $r\neq q$. Using this in 
equation \ref{eq:correlation}, we obtain that the correlation is
equal to $\frac{1}{2}$ (for $r\neq q$ and in the limit of large
$n$).

Recall that the problem Hamiltonian has been fixed and
therefore the  $\{d_{r i}\}$ are set. We have argued that for large n the differences $\left\{ \tilde{e}_{r}^{(2)}-\tilde{e}_{0}^{(2)}\right\} $
have approximately a joint normal distribution with pairwise correlations
equal to $\frac{1}{2}.$ The probability that a random choice of the $\{c_i\}$ make the algorithm succeed is \begin{eqnarray*}
\text{Pr}\left[\tilde{e}_r^{(2)}-\tilde{e}_0^{(2)}>0\text{ , \ensuremath{r=1,...,k-1}}\right]& = &\text{Pr}\left[\sum_{i=1}^{n}c_{i}^{2}d_{r i}>0\text{ , \ensuremath{r=1,...,k-1}}\right]\\
& = & \text{Pr}\left[\frac{\sum_{i=1}^{n}c_{i}^{2}d_{r i}-\frac{5}{4}\sum_{i=1}^{n}d_{r i}}{\sqrt{\sum_{i=1}^{n}\left(d_{r i}\right)^{2}}}>\frac{-\frac{5}{4}\sum_{i=1}^{n}d_{r i}}{\sqrt{\sum_{i=1}^{n}\left(d_{r i}\right)^{2}}}\text{,  \ensuremath{r=1,...,k-1}}\right].\end{eqnarray*}
 Note that in the last line we have subtracted the mean and divided by the standard deviation. Using a standard trick we can rewrite the above probability as (in the limit of large n), 
\begin{equation}
\label{prob_w}
\text{Pr}\left[w_{r}-w_0>\frac{-\frac{5}{4}\sum_{i=1}^{n}d_{r i}}{\sqrt{\sum_{i=1}^{n}\left(d_{r i}\right)^{2}}}\text{, $r=1,...,k-1$}\right],
\end{equation}
where $w_0$ and $\{w_{r}\}$ are independent normal random variables with
mean zero and variance $\frac{1}{2}$. This follows from the fact that the joint normal distribution is specified by its marginal distributions and pairwise correlations. Essentially $w_r$ and $w_0$ are the (normalized) changes in the quantities $e_r^{(2)}$ and $e_0^{(2)}$ provided by the randomization. 

At this point, in order to evaluate equation \ref{prob_w} we need to understand the magnitude of the RHS of the inequality. Consider typical values for the sum $\sum_{i=1}^{n}d_{r i}$. As in section \ref{sec:Problematic-Instances} we assume that this sum is Gaussian in the limit of large $n$ with mean $0$ and standard deviation approximated by $\sqrt{\sum_{i=1}^{n} d_{ri}^2}$.

(In the argument below, we use some properties of the Gaussian distribution
which follow from the inequalities \begin{eqnarray*}
\int_{a}^{\infty}e^{-x^{2}}dx & \leq & \int_{a}^{\infty}\frac{x}{a}e^{-x^{2}}dx=\frac{1}{2a}e^{-a^{2}}\\
\int_{a}^{\infty}e^{-x^{2}}dx & \geq & \int_{a}^{a+1}e^{-x^{2}}dx\geq e^{-(a+1)^{2}}.)\end{eqnarray*}

For the $k-1$ approximately Gaussian random variables $\frac{-\sum_{i=1}^{n}d_{r i}}{\sqrt{\sum_{i=1}^{n}\left(d_{r i}\right)^{2}}}$
with mean $0$ and variance $1$, for some constant $B$ it will hold with high probability
(in the limit of large $n$) that \[
\frac{-\frac{5}{4}\sum_{i=1}^{n}d_{r i}}{\sqrt{\sum_{i=1}^{n}\left(d_{r i}\right)^{2}}}<B\sqrt{\text{log}k}\text{ for \ensuremath{r=1,...,k-1}. }\]
Now the random variable $w_0$ satisfies $w_0<-2B\sqrt{\log k}$ with
probability which is only polynomially small in $k$. For each $w_{r}$
the probability that $w_{r}< -B\sqrt{\text{log}k}$ is polynomially
small in $k$ and we use the bound\[
\text{Pr}\left[w_r-w_{0}\geq B\sqrt{\text{log}k}\text{, \ensuremath{r=1,...,k-1}}\right]\geq\text{Pr}\left[w_0<-2B\sqrt{\log k}\right]\cdot\prod_{r=1}^{k-1}\left(1-\text{Pr}\left[w_{r}< -B\sqrt{\text{log}k}\right]\right).\]
The first term on the RHS is only polynomially small, and each of the $k-1$ terms in the product are at least $1-\frac{1}{k^2}$ if $B$ is large enough. So the probability that the adiabatic algorithm succeeds is at worst polynomially small in $k$. By repeating the randomization a polynomial number of times we expect to succeed with high probability.

\section{Quantum Monte Carlo and Numerical Results\label{sec:Quantum-Monte-Carlo}}

\subsection*{Continuous Imaginary Time Quantum Monte Carlo}

This section is a review of continous imaginary time Quantum Monte
Carlo (which is a classical path integral simulation technique for
extracting properties of quantum systems \cite{prokofev-1996-64}).
In particular we will show how this method can be used to compute
thermal expectation values of Hermitian operators at inverse temperature
$\beta$. We start with a Hamiltonian H which we write as \[
H=H_{0}+V\]
 where $H_{0}$ is diagonal in some known basis $\{|z\rangle\}$,
and V is purely off diagonal in this basis. We require that all the
nonzero matrix elements of $V$ are negative. For the Hamiltonian
$H(s)=(1-s)H_{B}+sH_{P}$ with $H_{B}$ as in equation \ref{eq:Hb},
we have \begin{eqnarray*}
H_{0} & = & sH_{P}+\frac{(1-s)n}{2}\\
V & = & -(1-s)\sum_{i=1}^{n}\frac{\sigma_{x}^{i}}{2}\,.\end{eqnarray*}
 Here we include the factor of $1-s$ in the definition of $V$ since
we are not doing perturbation theory in this quantity.

The partition function can be expanded using the Dyson series as follows
\begin{align*}
 & Tr\left[e^{-\beta H}\right]=Tr\bigg[e^{-\beta H_{0}}\sum_{m=0}^{\infty}(-1)^{m}\int_{t_{m}=0}^{\beta}dt_{m}\int_{t_{m-1}=0}^{t_{m}}dt_{m-1}...\int_{t1_{1}=0}^{t_{2}}dt_{1}V_{I}(t_{m})...V_{I}(t_{1})\bigg]\,.\end{align*}
 Here we use the notation $V_{I}(t)=e^{tH_{0}}Ve^{-tH_{0}}$. We can
then insert complete sets of states and take the trace to obtain the
path integral \begin{align}
Tr\left[e^{-\beta H}\right]=\nonumber \\
 & \sum_{m=0}^{\infty}\sum_{\{z_{1},...,z_{m}\}}\bigg[(-1)^{m}\langle z_{1}|V|z_{m}\rangle\langle z_{m}|V|z_{m-1}\rangle...\langle z_{2}|V|z_{1}\rangle\nonumber \\
 & \int_{t_{m}=0}^{\beta}dt_{m}\int_{t_{m-1}=0}^{t_{m}}dt_{m-1}...\int_{t_{1}=0}^{t_{2}}dt_{1}e^{-\int_{t=0}^{\beta}\mathcal{H}_{0}(z(t))dt}\bigg]\,.\label{eq:pathint}\end{align}
 In this formula we have used the notation $\mathcal{H}_{0}(z(t))=\langle z(t)|H_{0}|z(t)\rangle$,
where the function $z(t)$ is defined by\begin{eqnarray*}
z(t) & = & \begin{cases}
z_{1}, & 0\leq t<t_{1}\\
z_{2}, & t_{1}\leq t<t_{2}\\
\;\vdots\\
z_{m}, & t_{m-1}\leq t<t_{m}\\
z_{1}, & t_{m}\leq t\leq\beta\,.\end{cases}\end{eqnarray*}
 So in particular \[
\int_{t=0}^{\beta}\mathcal{H}_{0}(z(t))dt=\langle z_{1}|H_{0}|z_{1}\rangle(t_{1}+\beta-t_{m})+\langle z_{2}|H_{0}|z_{2}\rangle(t_{2}-t_{1})+...\langle z_{m}|H_{0}|z_{m}\rangle(t_{m}-t_{m-1})\,.\]
 We view the function $z(t)$ as a path in imaginary time which begins
at $t=0$ and ends at $t=\beta.$ Then equation \ref{eq:pathint}
is a sum over paths, where every path is assigned a positive weight
according to a measure $\tilde{\rho}$

\[
\tilde{\rho}(P)=(-1)^{m}\langle z_{1}|V|z_{m}\rangle\langle z_{m}|V|z_{m-1}\rangle...\langle z_{2}|V|z_{1}\rangle dt_{1}...dt_{m}e^{-\int_{t=0}^{\beta}\mathcal{H}_{0}(z(t))dt}\,.\]
 Note that the fact that $\tilde{\rho}$ is positive semidefinite
follows from our assumption that all matrix elements of $V$ are negative.
We write \begin{equation}
\rho=\frac{\tilde{\rho}}{Z(\beta)}\label{eq:rho}\end{equation}
 for the normalized distribution over paths.

We now discuss how one can obtain properties of the quantum system
as expectation values with respect to the classical measure $\rho.$
We are interested in ground state properties, but in practice we select
a large value of $\beta$ and compute thermal expectation values at
this inverse temperature. If $\beta$ is sufficiently large then these
expectation values will agree with the corresponding expectation values
in the ground state. It is straightforward to show that for any operator
$A$ which is diagonal in the $|z\rangle$ basis, the thermal expectation
value can be written as\begin{equation}
\frac{Tr[Ae^{-\beta H}]}{Tr[e^{-\beta H}]}=\langle\frac{1}{\beta}\int_{0}^{\beta}\mathcal{A}(z(t))dt\rangle_{\rho}\label{eq:diagonaest}\end{equation}
 where $\mathcal{A}(z(t))=\langle z(t)|A|z(t)\rangle.$ The notation
$\langle\rangle_{\rho}$ means average with respect to the classical
probability distribution $\rho$ over paths.

This immediately allows us to estimate quantities such as the diagonal
part of the Hamiltonian\begin{eqnarray}
\frac{Tr[H_{0}e^{-\beta H}]}{Tr[e^{-\beta H}]} & = & \langle\frac{1}{\beta}\int_{0}^{\beta}\mathcal{H}_{0}(z(t))dt\rangle_{\rho}\,.\label{eq:H0}\end{eqnarray}
 Another quantity that we find useful in our study is the Hamming
weight operator defined by\begin{equation}
W=\sum_{i=1}^{n}\left(\frac{1-\sigma_{z}^{i}}{2}\right)\,.\label{eq:Weight}\end{equation}
 which can also be estimated using equation \ref{eq:diagonaest}.
In order to estimate the thermal expectation value of the full Hamiltonian
$H_{0}+V$ we use equation \ref{eq:H0} as well as the expression\begin{equation}
\frac{Tr[Ve^{-\beta H}]}{Tr[e^{-\beta H}]}=-\langle\frac{m}{\beta}\rangle_{\rho}\label{eq:V}\end{equation}
 where $m$ is the number of transitions in the path. (Equation \ref{eq:V}
is not as simple to derive as equation \ref{eq:diagonaest}.) So by
generating paths from the distribution $\rho$ and then computing
averages with respect to $\rho$, we can evaluate the thermal expectation
value of the energy at a given inverse temperature $\beta$ and by
taking $\beta$ sufficiently large we can approximate the ground state
energy.

Generating paths from the distribution $\rho$ is itself a challenging
task. We use a modified version of the heat bath algorithm of Krzakala
et al \cite{krzakala-2008-78} which we describe in the appendix.
As with other Quantum Monte Carlo methods, this algorithm is a Markov
Chain Monte Carlo method. In order to sample from the distribution
$\rho$, one defines a Markov Chain over the state space consisting
of all paths, where the limiting distribution of the chain is $\rho$.
To obtain samples from $\rho$ one starts in an initial path $z_{init}(t)$
and then applies some number $N_{equil}$ of iterations of the Markov
Chain. If $N_{equil}$ is sufficiently large then the distribution
of the paths found by further iteration will be close to $\rho.$
We use these subsequent paths to compute averages with respect to
$\rho$.

\subsection*{Equilibration of the Quantum Monte Carlo and Identification of Level
Crossings}

The discussion in the previous section demonstrates how one can estimate
expectation values of various quantities in the ground state of a
quantum system. The method is {}``continuous time'' so there is
no discretization error and for fixed $\beta$ quantities can be arbitrarily
well approximated with enough statistics. We use the Quantum Monte
Carlo method in a nonstandard way in order to be able to study the
lowest two eigenstates of our Hamiltonian (as opposed to just the
ground state).

As described in the previous section, the standard procedure for generating
configurations is to equilibrate to the distribution $\rho$ from
some initial path $z_{init}(t)$ (we call this the seeded path) by
applying the Monte Carlo update $N_{equil}$ times. In order to ensure
that one has equilibrated after $N_{equil}$ Monte Carlo updates,
one could for example do simulations with two or more different initial
paths (seeded paths) and check that the values of Monte Carlo observables
appear to converge to the same seed independent values.

For each instance we consider, we run two different Monte Carlo simulations
that are seeded with two different paths $z_{init}^{0}(t)$ and $z_{init}^{1}(t)$.
These seeds are paths with no flips in them, corresponding to the
two states $000...0$ and $111...1$,\begin{eqnarray*}
z_{init}^{0}(t) & = & 000...0\text{ for all }t\in[0,\beta]\\
z_{init}^{1}(t) & = & 111...1\text{ for all }t\in[0,\beta]\,.\end{eqnarray*}
 With each seed, we run the Monte Carlo simulation for some total
number $N_{total}$ of Monte Carlo sweeps, taking data every $k$
sweeps. Here a single sweep is defined to be $n$ iterations of our
Monte Carlo update rule as described in the appendix (where $n$ is
the number of spins). We then remove the first $N_{equil}$ Monte
Carlo samples, and use the remaining Monte Carlo samples to estimate
the thermal averages \begin{eqnarray*}
\langle H\rangle & = & \frac{Tr[He^{-\beta H}]}{Tr[e^{-\beta H}]}\\
\langle W\rangle & = & \frac{Tr[We^{-\beta H}]}{Tr[e^{-\beta H}]}\end{eqnarray*}
 using equations \ref{eq:H0},\ref{eq:V} and \ref{eq:Weight}.

\begin{figure}[p]
 \centering{}\includegraphics[scale=0.95]{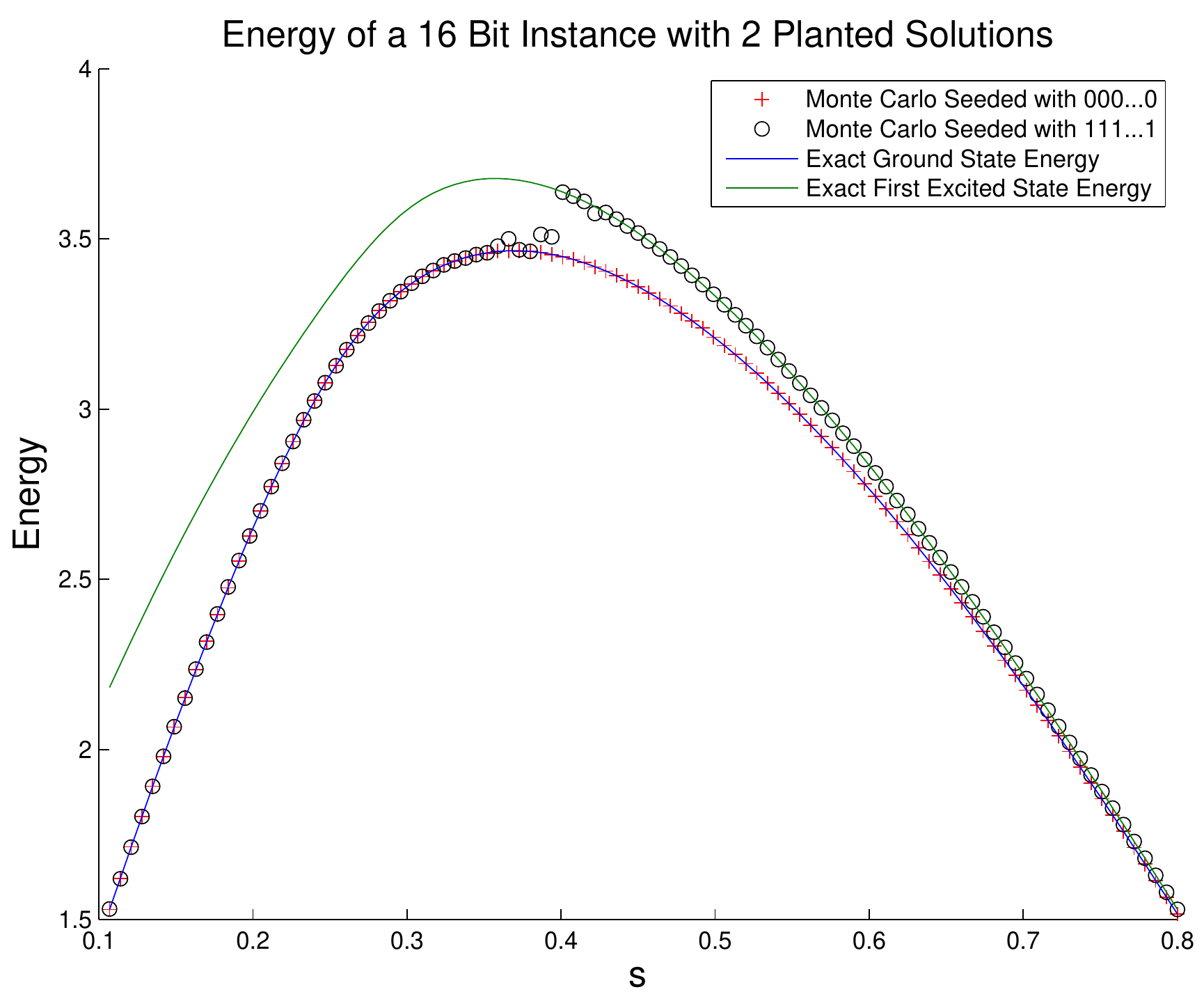}

\caption{The discontinuity in the circle data that occurs near $s=0.4$ is
a Monte Carlo effect that we understand. As can be seen from the exact
numerical diagonalization there is no true discontinuity in either
the ground state energy or the first excited state energy. For $s$
greater than $0.4$ the Monte Carlo simulation is in a metastable
equilibrium that corresponds to the first excited state. The true
ground state energy at each $s$ is always the lower of the circle
and the cross at that value of $s.$}

\label{Flo:16_nobitfix} 
\end{figure}

\begin{figure}[p]

\begin{centering}
\includegraphics[scale=0.95]{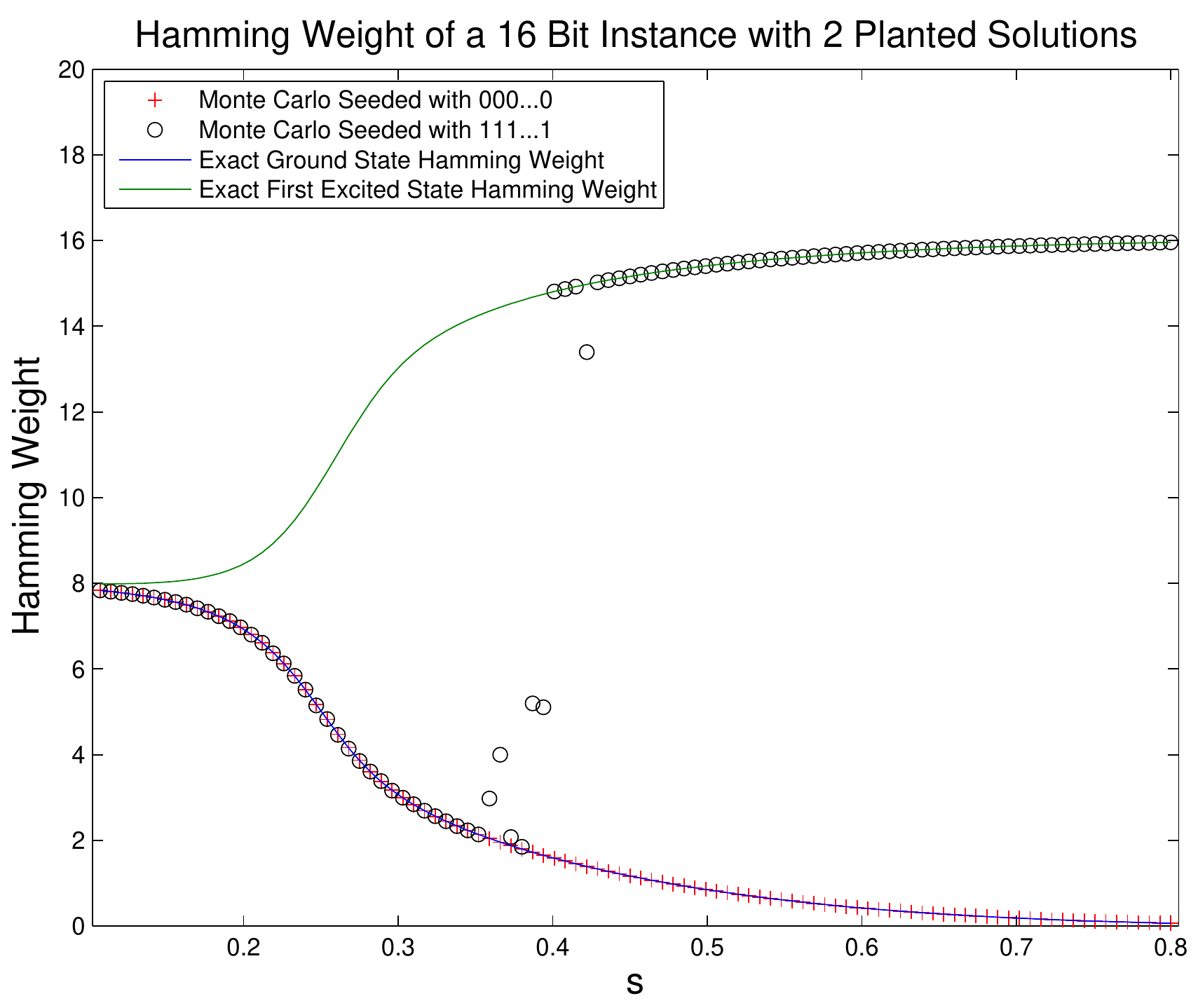} 
\par\end{centering}

\caption{Together with figure \ref{Flo:16_nobitfix}, we see that the discontinuity
in the circles appears in data for both the energy and the Hamming
weight. This is not indicative of a phase transition in the physical
system (as evidenced by the smooth curves computed by exact numerical
diagonalization), but is purely a result of the way in which we use
the Monte Carlo method.}

\label{Flo:16_mag_nobitfix} 
\end{figure}

In order to convince the reader that the Monte Carlo algorithm works
correctly, and that we can obtain information about the first excited
state, we show data at 16 bits where exact numerical diagonalization
is possible. In figures \ref{Flo:16_nobitfix} and \ref{Flo:16_mag_nobitfix}
we show the result of this procedure for a 3SAT instance with 16 bits,
with problem Hamiltonian $H_{P}^{'}$ (before adding the penalty term
$h$, so that the levels are degenerate at $s=1$ ). This instance
has 122 clauses. The inverse temperature is $\beta=150$. The total
number of Monte Carlo sweeps at each value of $s$ is $N_{total}=200000$,
with data taken after every fifth sweep (this gave us $40000$ data
samples). We use $N_{equil}=2500$ samples solely for equilibration
at each value of $s$. Note that the Monte Carlo simulations with
the two different seeds (corresponding to the circles and crosses
in the figures) only agree for values of $s$ less than roughly $0.4$.
We interpret this to mean that at these values of $s$ the Quantum
Monte Carlo has equilibrated to the proper limiting distribution $\rho$
regardless of the seed. As $s$ increases past $0.4$, the two simulations
abruptly begin to give different results. In this case the simulation
seeded with $z_{init}^{1}(t)$ finds a metastable equilibrium of the
Markov Chain (i.e not the limiting distribution $\rho$), which corresponds
in this case to the first excited state of the Hamiltonian. For $s$
above $0.4$ we can see from the comparison with exact diagonalization
that the two differently seeded values allow us to compute the energies
(figure \ref{Flo:16_nobitfix}) and Hamming weights (figure \ref{Flo:16_mag_nobitfix})
of the lowest two energy levels of our Hamiltonian.

What is happening here is that for large $s$, the quantum system
can be thought of as consisting of two disconnected sectors. One sector
consists of states in the z basis with low Hamming weight and the
other with Hamming weight near $n$. The Quantum Monte Carlo {}``equilibrates''
in each sector depending on the initial seed as can be seen by the
smooth data in each sector. (Note that data is taken independently
at each value of s.) The lower of the cross and circle at each value
of $s$ in figure \ref{Flo:16_nobitfix} is the ground state energy.
Of course if the classical algorithm ran for long enough then the
circles in figure \ref{Flo:16_nobitfix} would lie on the crosses
for every value of $s$.

\begin{figure}[p]

\begin{centering}
\includegraphics[scale=0.95]{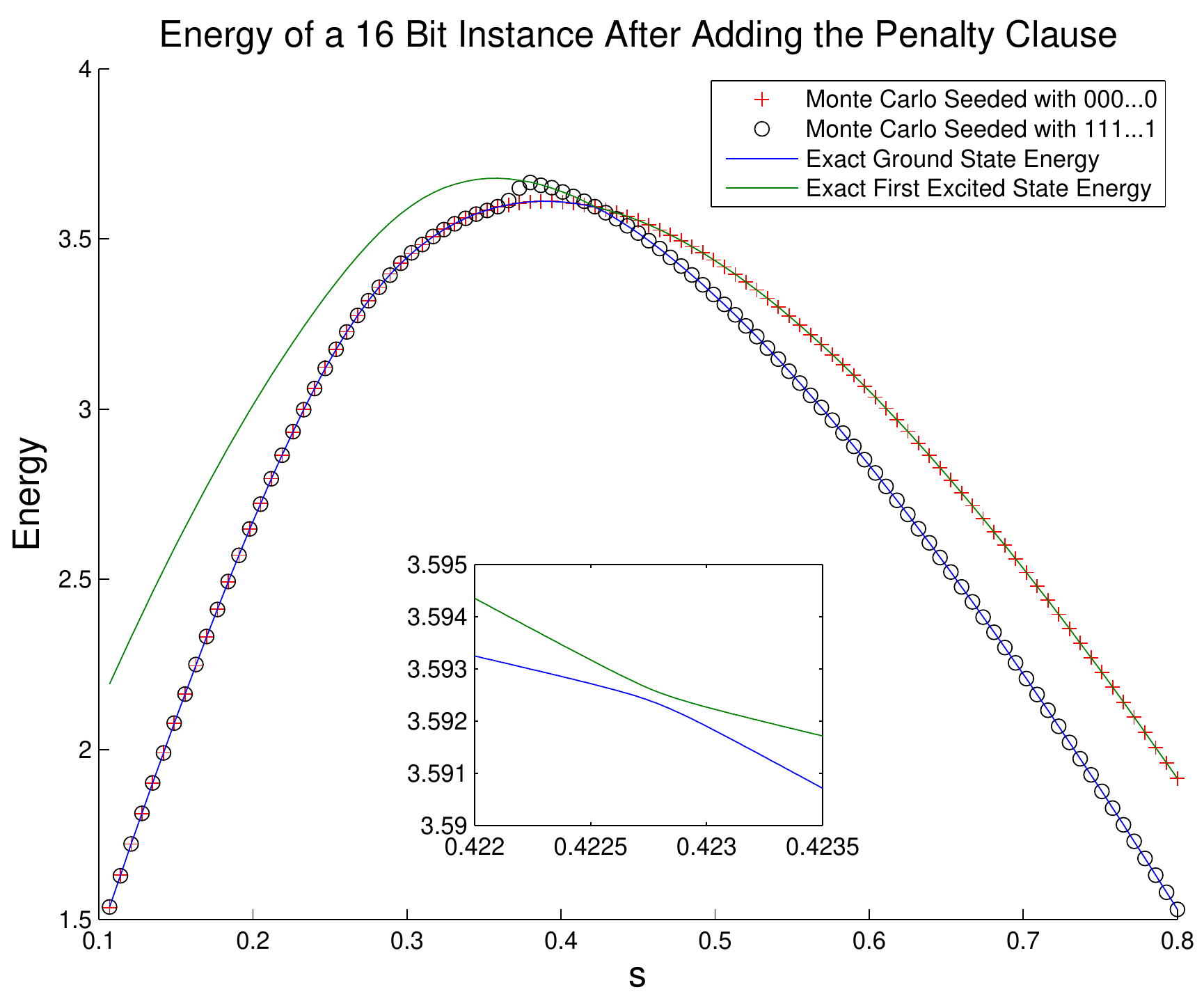} 
\par\end{centering}

\caption{After adding the penalty clause we see that the energy levels have
an avoided crossing at $s\approx0.423$. The inset shows exact numerical
diagonalization near the avoided crossing where we can resolve a tiny
gap. The ground state energy is well approximated by the lower of
the circle and cross at each value of $s$.}

\label{Flo:16_cross} 
\end{figure}

\begin{figure}[p]

\begin{centering}
\includegraphics[scale=0.95]{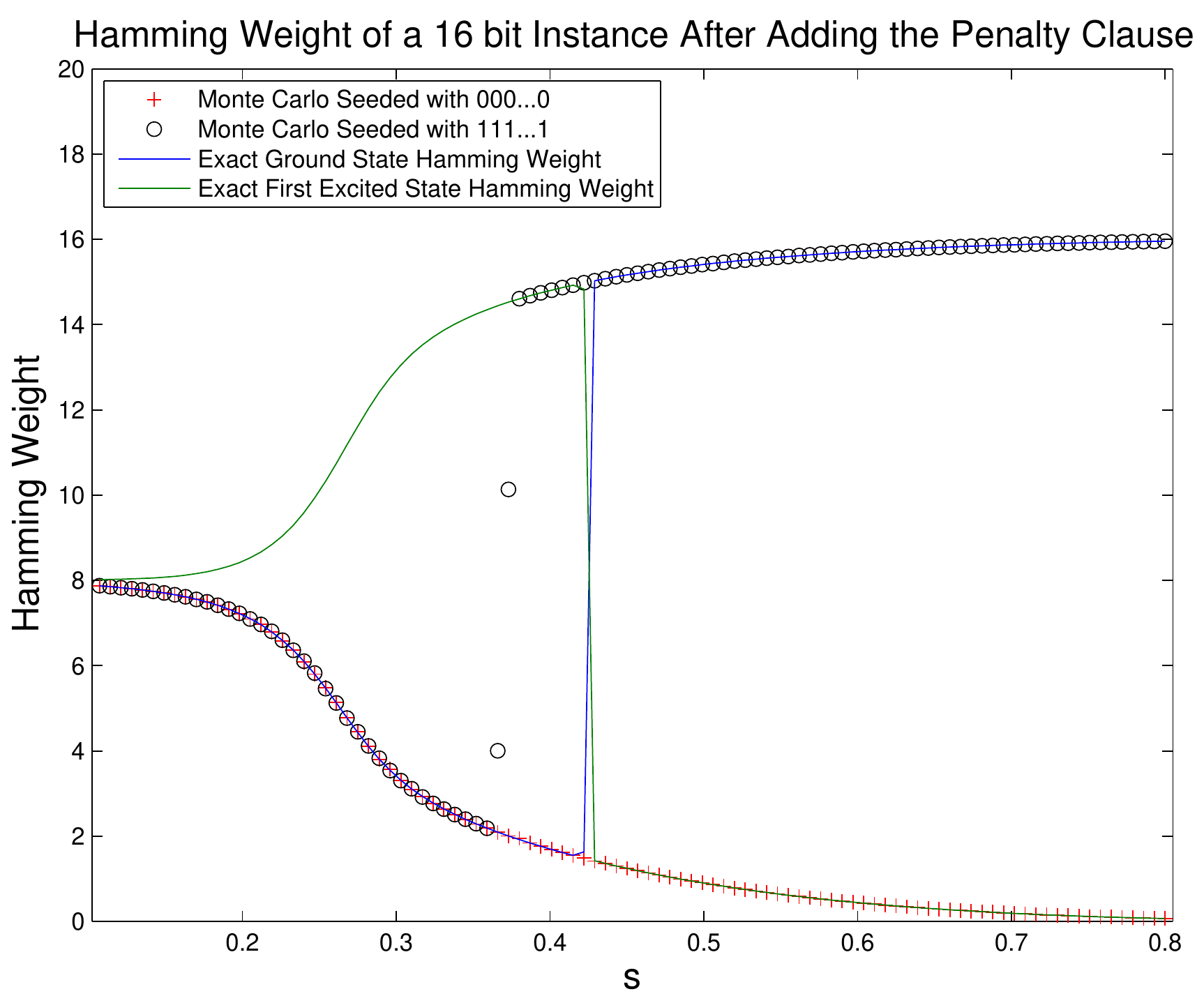} 
\par\end{centering}

\caption{In this figure there is a phase transition which occurs near $s\approx0.423$.
We see from the exact numerical diagonalization that the Hamming weights
of the first excited state and ground state undergo abrupt transitions
at the point where there is a tiny avoided crossing in figure \ref{Flo:16_cross}.
There is also a jump in the Monte Carlo data plotted with circles
that occurs before the avoided crossing: this is a Monte Carlo effect
as discussed earlier and has no physical significance. If we look
at the Monte Carlo data in figure \ref{Flo:16_cross} we conclude
that below $s\approx0.423$ the crosses represent the ground state
and after this point the circles represent the ground state. From
the current figure, along with figure \ref{Flo:16_cross}, we conclude
that the Hamming weight of the ground state jumps abruptly at $s^{\star}\approx0.423$.}

\label{Flo:16_cross_mag} 
\end{figure}

In figures \ref{Flo:16_cross} and \ref{Flo:16_cross_mag} we show
data taken for the same instance after the penalty clause is added
that removes the degeneracy at $s=1$. In this case the two levels
have an avoided crossing as expected and which can be seen by exact
numerical diagonalization in figure \ref{Flo:16_cross}. However in
the Monte Carlo simulation with two seeds there are two essentially
disconnected sectors and the two levels being tracked do cross. When
we see this behaviour in the Monte Carlo simulation, we interpret
it as compelling evidence that there is a tiny gap in the actual system,
which occurs where the curves cross. In figure \ref{Flo:16_cross_mag}
first look at the curves coming from the exact numerical diagonalization.
The Hamming weight of the ground state decreases and then abruptly
rises at the location of the minimum gap. The Hamming weight of the
first excited state also undergoes an abrupt change at the location
of the minimum gap. Not surprisingly the exact diagonalization clearly
shows the behaviour we expect with the manufactured tiny gap. Now
look at the Monte Carlo data which is interpreted by first looking
at figure \ref{Flo:16_cross}. In figure \ref{Flo:16_cross} the true
ground state is always the lower of the circles and the crosses. For
$s$ below $s^{\star}\approx0.423$ it is the crosses and for $s$
larger than $s^{\star}$ it is the circles. Accordingly in figure
\ref{Flo:16_cross_mag} the Hamming weight of the ground state is
tracked by the crosses for $s$ below $s^{\star}$ and by the circles
for $s$ above $s^{\star}$. We can therefore conclude, based only
on the Monte Carlo data, that the Hamming weight of the ground state
changes abruptly. At higher bit number we do not have exact numerical
diagonalization available but we can still use the Quantum Monte Carlo
to extract key information.

\subsection*{Randomizing the Beginning Hamiltonian}

For the 16 spin Hamiltonian depicted above, we now consider randomizing
the beginning Hamiltonian $H_{B}$ as described in section \ref{sec:Fixing-the-Problem}.
In order to minimize the number of times we run the Monte Carlo algorithm
(this will be more of an issue at high bit number where simulations
are very time consuming), we generated many different sets of coefficients
$\{c_{i}\}$ and calculated the differences $\tilde{e}_{U}^{(2)}-\tilde{e}_{L}^{(2)}$
with fixed problem Hamiltonian $H_{P}^{\prime}$ and beginning Hamiltonians
\[
\tilde{H}_{B}=\sum_{i=1}^{n}c_{i}\left(\frac{1-\sigma_{x}^{i}}{2}\right)\,.\]

\begin{figure}[p]

\begin{centering}
\includegraphics[scale=0.95]{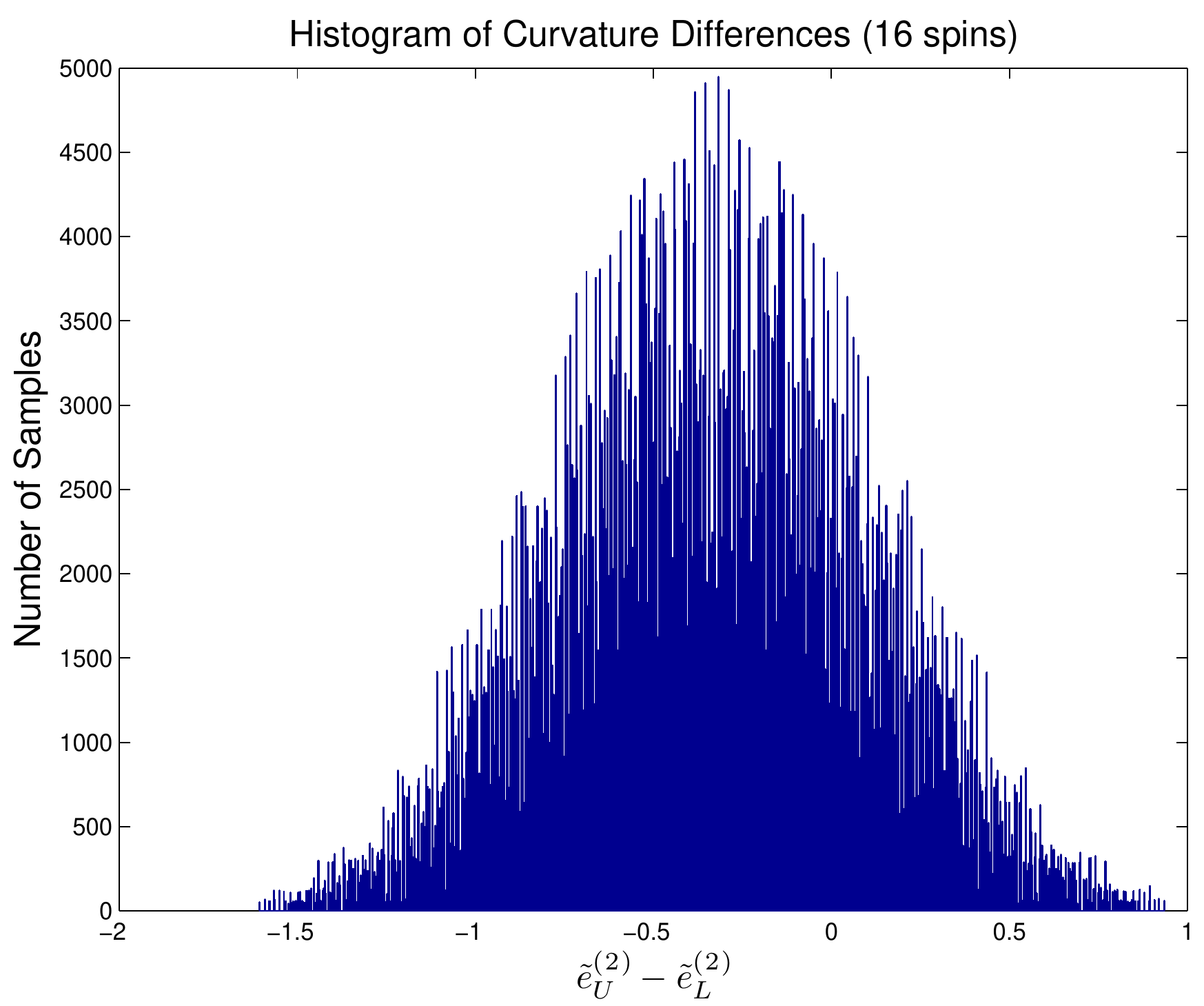} 
\par\end{centering}

\caption{The histogram of $\tilde{e}_{U}^{(2)}-\tilde{e}_{L}^{(2)}$ for 1
million different choices of coefficients $c_{i}$ shows a substantial
tail for which $\tilde{e}_{U}^{(2)}-\tilde{e}_{L}^{(2)}>0$. These
sets of coefficients correspond to beginning Hamiltonians $\tilde{H}_{B}$
for which we expect the small gap in figure \ref{Flo:16_cross} to
be removed. }

\label{Flo:histn16}
\end{figure}

\begin{figure}[p]

\begin{centering}
\includegraphics[scale=0.95]{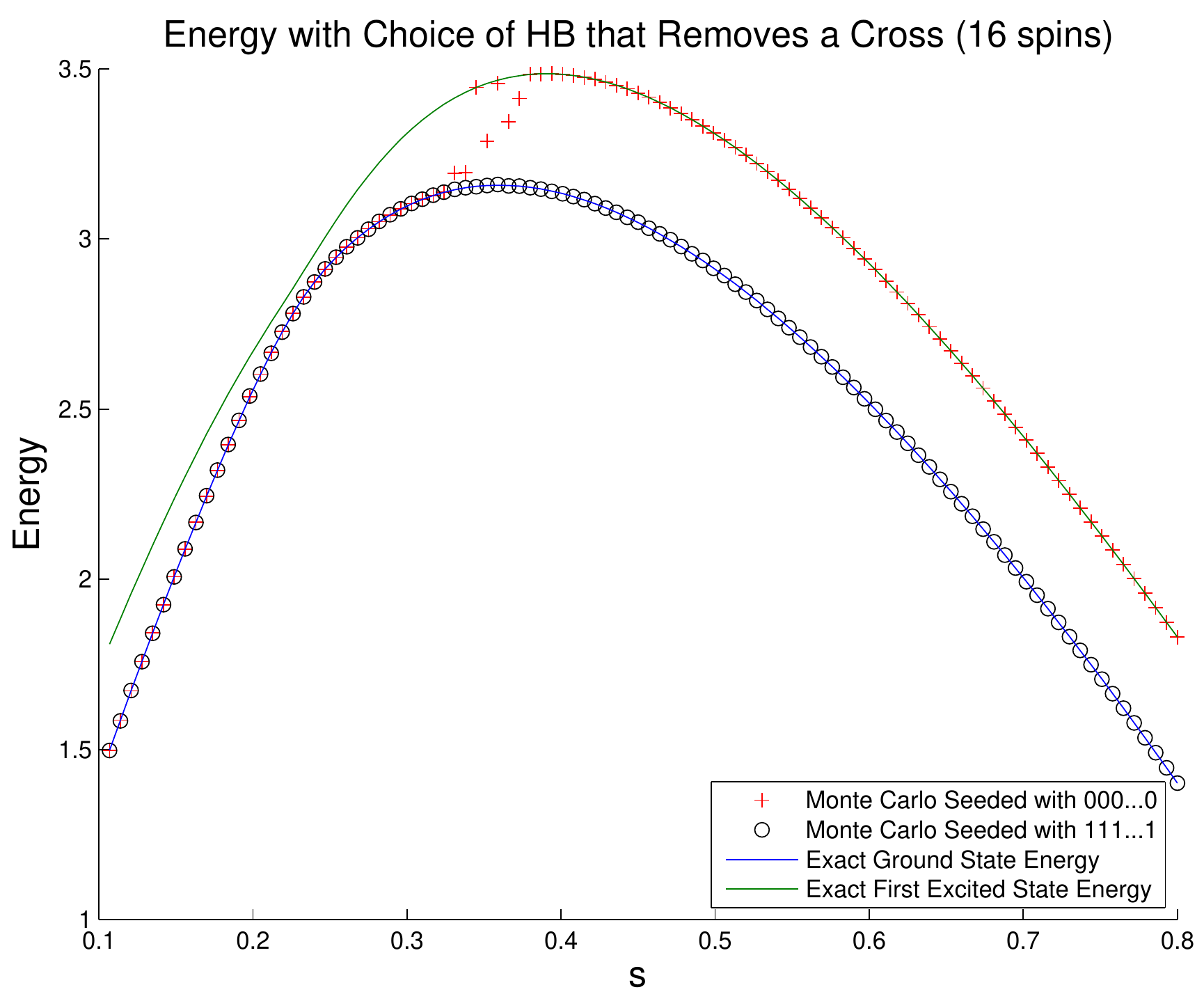} 
\par\end{centering}

\caption{A random beginning Hamiltonian removes the crossing seen in figure
\ref{Flo:16_cross}. The problem Hamiltonian is the same as that in
figure \ref{Flo:16_cross}. The circles are always below (or equal
to) the crosses for all values of $s$. This means that the circles
track the ground state for all $s$ and we see no sign of a small
gap in the Monte Carlo data. This is consistent with the displayed
exact numerical diagonalization.}

\label{Flo:16_hb1}
\end{figure}

\begin{figure}[ph]

\begin{centering}
\includegraphics[scale=0.95]{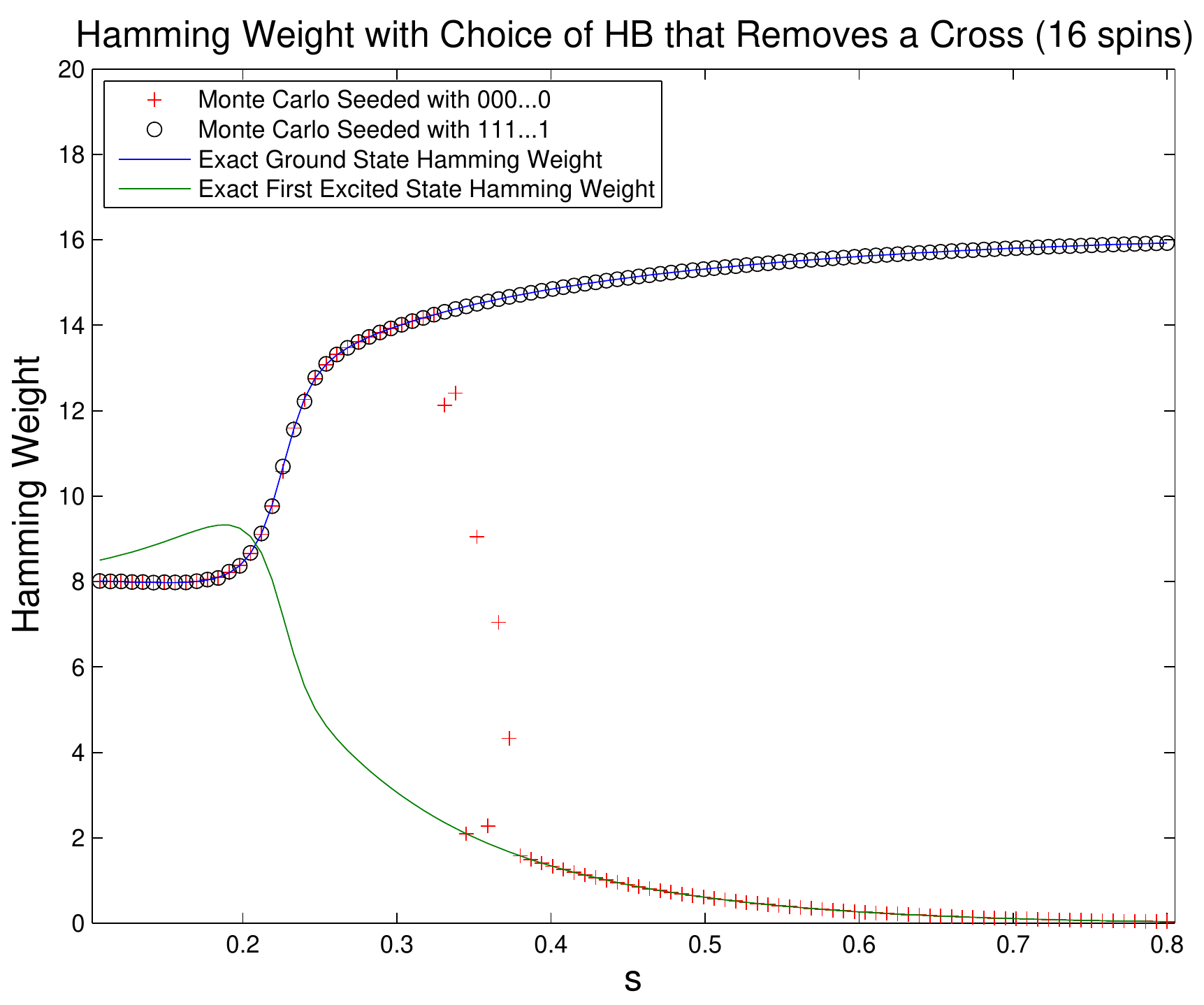} 
\par\end{centering}

\caption{From figure \ref{Flo:16_hb1} we see that the ground state of the
Hamiltonian corresponds to the circles for all values of $s$ and
the Hamming weight of the circles here goes smoothly to the Hamming
weight of the unique satisfying assignment. (The jump in the Monte
Carlo data corresponding to the crosses is due to the Monte Carlo
effect discussed earlier.)}

\label{Flo:16_mag_hb1}
\end{figure}

According to our discussion in section \ref{sec:Fixing-the-Problem},
we expect the avoided crossing near $s=1$ to be removed for choices
of coefficients $c_{i}$ such that the difference $\tilde{e}_{U}^{(2)}-\tilde{e}_{L}^{(2)}>0$
. We made a histogram of the $\tilde{e}_{U}^{(2)}-\tilde{e}_{L}^{(2)}$
(shown in figure \ref{Flo:histn16}) and randomly chose three sets
of coefficients $\{c_{i}\}$ such that $\tilde{e}_{U}^{(2)}-\tilde{e}_{L}^{(2)}>\frac{1}{2}$
for each. Our analysis predicts that in these cases the crossing will
be removed. As expected, each of these three sets of coefficients
resulted in an adiabatic Hamiltonian with the small gap removed, although
in one there was another small gap at a smaller value of $s$. We
plot the Monte Carlo data for one of these sets in figures \ref{Flo:16_hb1}
and \ref{Flo:16_mag_hb1}.

In section \ref{sec:Problematic-Instances} we argued that we could
manufacture a small gap in a 3SAT instance as sketched in figure \ref{Flo:after}.
Our 16 bit instance is a concrete example of this and the tiny gap
can be seen in figure \ref{Flo:16_cross}. In section \ref{sec:Fixing-the-Problem}
we argued that by randomizing the beginning Hamiltonian we should
be able to produce figure \ref{Flo:fixed}. This is what we see concretely
at 16 bits in figure \ref{Flo:16_hb1}. We now tell the same story
at 150 bits using only Monte Carlo data since exact numerical diagonalization
is not possible.

\subsection*{Data for an Instance of 3SAT with 150 bits}

In addition to validating our method at 16 bits, we studied 3SAT instances
with $25,75$ and $150$ bits using our Quantum Monte Carlo simulator.
The data from these simulations for the most part supported our arguments.
In this section we present Monte Carlo data taken for a double plant
instance of 3SAT with 150 spins and 1783 clauses. In this case the
inverse temperature $\beta=300$, and the total number of Monte Carlo
sweeps at each value of $s$ is $N_{total}=100000$, with data taken
every fifth sweep (giving $20000$ data samples). The first $2500$
data samples at each value of $s$ are removed for equilibration.
We ran our simulations on a 648 processor SiCortex computer cluster
in embarrassingly parallel fashion. We used a different processor
for each value of $s$ and each value of the seed. Data taken at the
lower values of $s$ took the longest to accumulate, in some cases
more than 10 days on a single processor for a single data point.

\begin{figure}[p]
 \begin{raggedright} \includegraphics[scale=0.95]{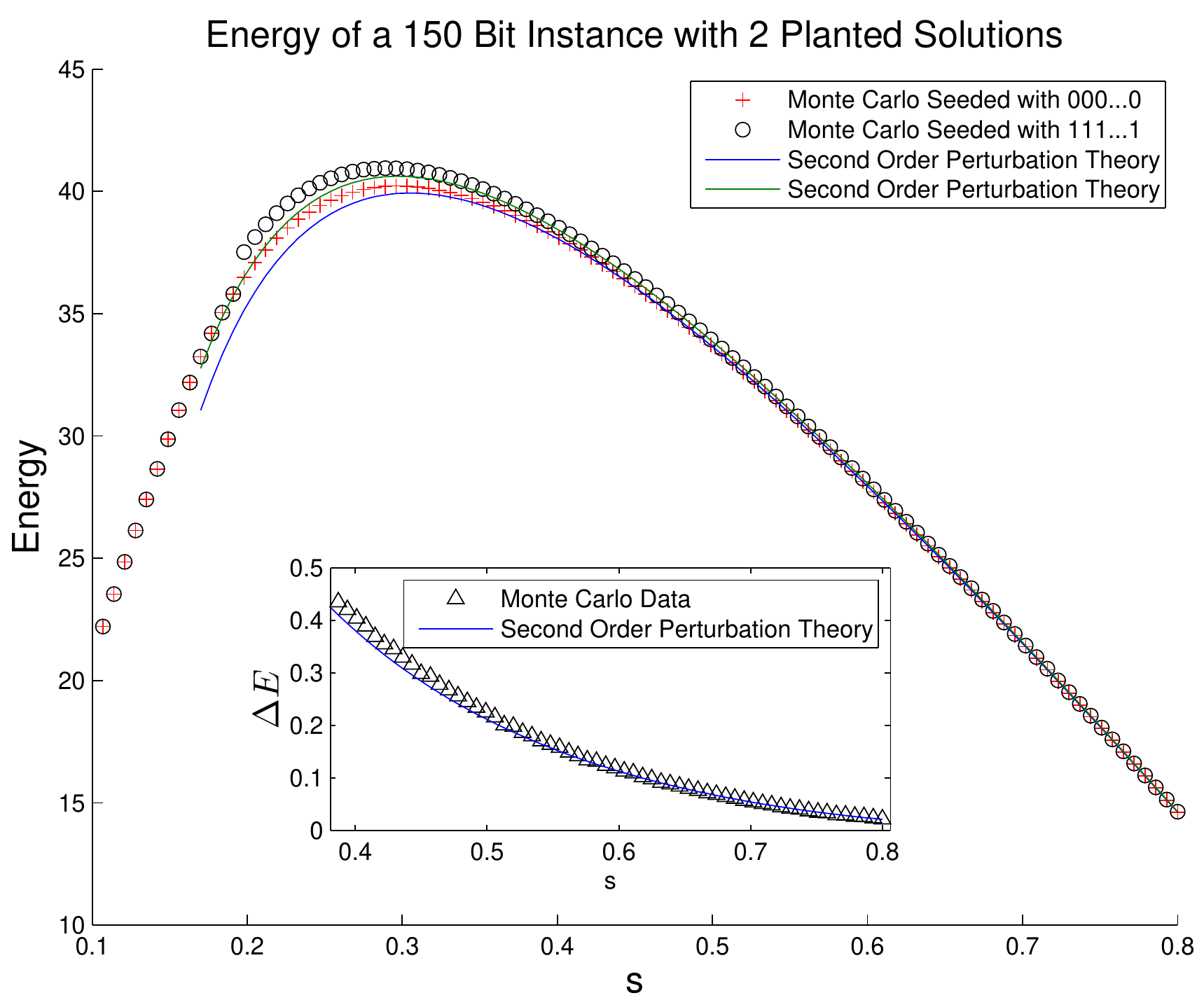} 

\end{raggedright}

\caption{The crosses, which represent the Monte Carlo data seeded with $111...1$,
are below (or equal to) the circle data for all values of $s$. (This
is seen more clearly in the inset which shows the positive difference
between the circle and cross values). We conclude that the crosses
track the ground state energy which is smoothly varying. The jump
in the circle data is a Monte Carlo effect and for $s$ above $0.2$
the circles track the first excited state. }

\label{Flo:150_nocross}
\end{figure}

\begin{figure}[p]
 \begin{raggedright} \includegraphics[scale=0.95]{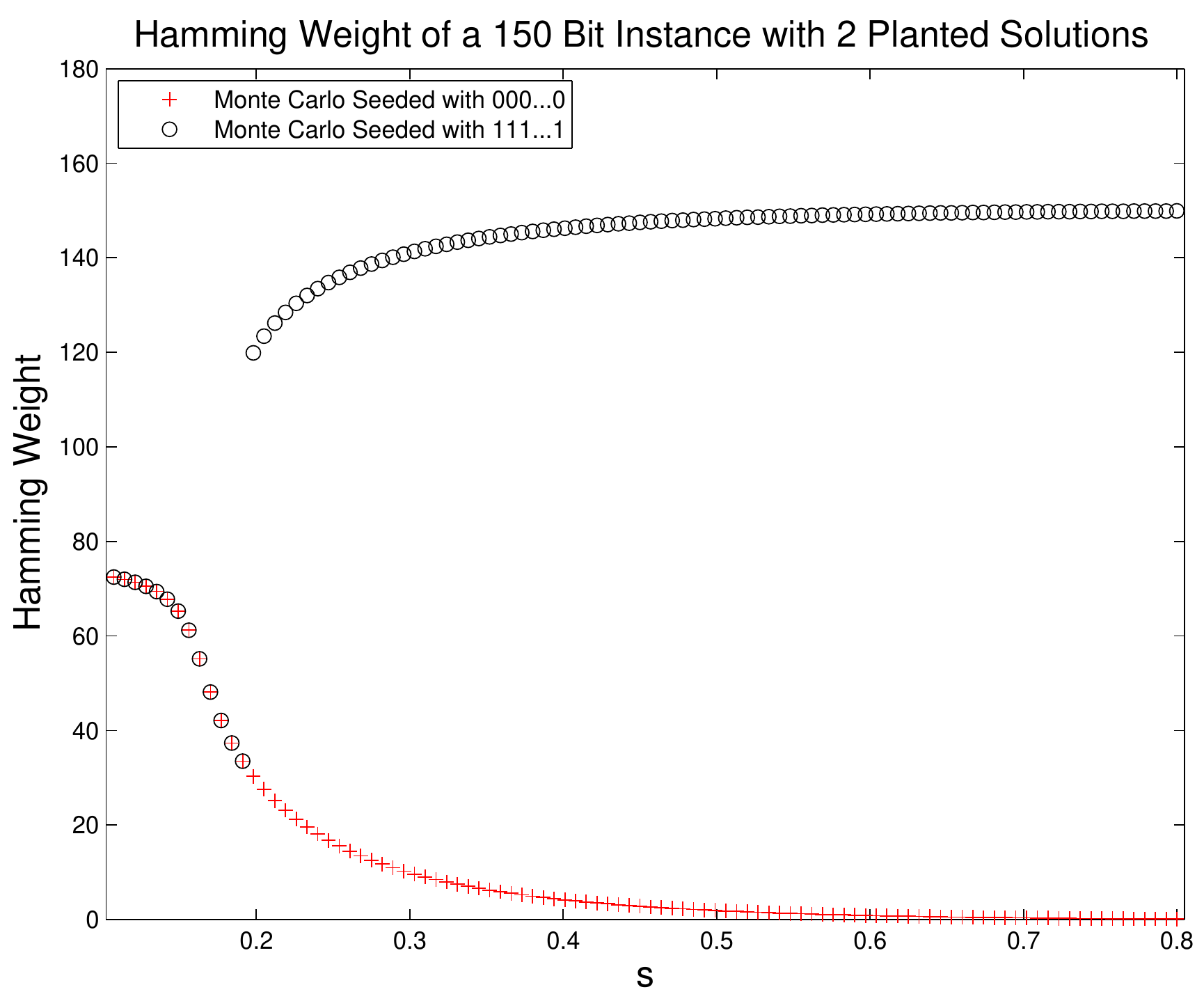} 

\end{raggedright}

\caption{Comparing with figure \ref{Flo:150_nocross} we see that the Hamming
weight for the first excited state and the ground state are continous
functions of $s$ . We only obtain data for the first excited state
for values of $s$ larger than $s\approx0.2$ .}

\label{Flo:150_mag_nocross}
\end{figure}

Figures \ref{Flo:150_nocross} and \ref{Flo:150_mag_nocross} show
the energy and the Hamming weight before the penalty clause is added,
running with two different seeds. In figure \ref{Flo:150_nocross}
for large values of $s$, the crosses are always below the circles
and track the ground state which ends at $000...0$ as can be seen
in figure \ref{Flo:150_mag_nocross}. We have also plotted the second
order perturbation theory energies for these two levels, expanding
around $s=1$. Note the good agreement between the second order perturbation
theory and the Monte Carlo data all the way down to $s=0.35$.

\begin{figure}[p]
 \begin{raggedright} \includegraphics[scale=0.95]{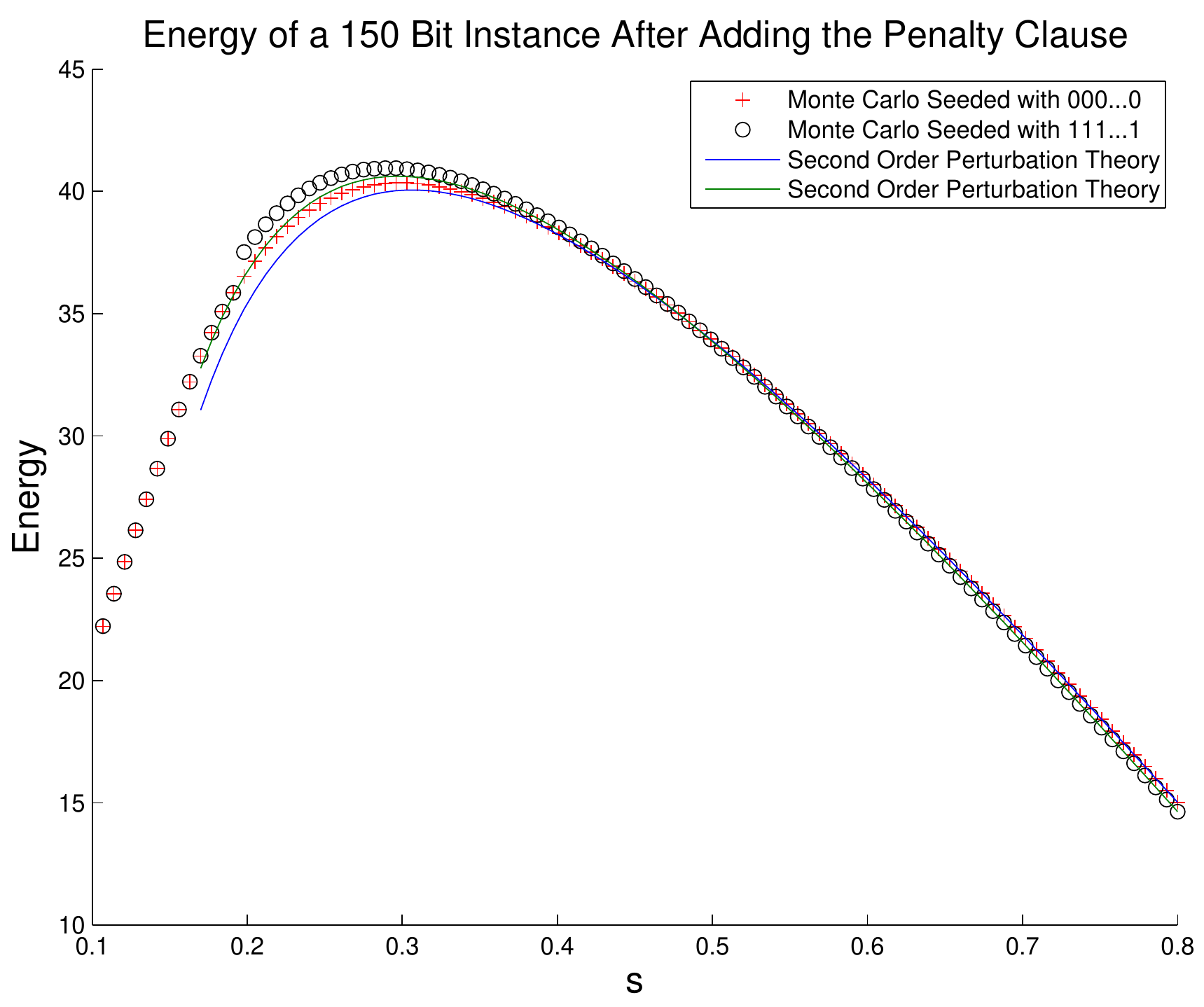} 

\end{raggedright}

\caption{Adding the penalty clause makes the cross data go above the circle
data at $s\approx0.49$. This is shown in more detail in figure \ref{Flo:150_expanded}
where we plot the energy difference between the first two levels as
a function of $s$. We interpret this to mean that the Hamiltonian
$H(s)$ has a tiny gap at $s^{\star}\approx0.49$.}

\label{Flo:150_cross}
\end{figure}

Since the lower curve corresponds to $000...0$, we penalize this
assignment by the addition of a single clause to form $H_{P}$ attempting
to manufacture a near crossing. In figure \ref{Flo:150_cross}, the
circle data is below the cross data for $s$ near $1$ but the two
curves cross near $s=.49$. This is seen clearly in figure \ref{Flo:150_expanded}
where the energy difference is plotted. From the Monte Carlo data
we conclude that $H(s)$ has a tiny gap. The location of the avoided
crossing is well predicted by second order perturbation theory as
can be seen in figure \ref{Flo:150_expanded}.

\begin{figure}
\includegraphics[scale=0.95]{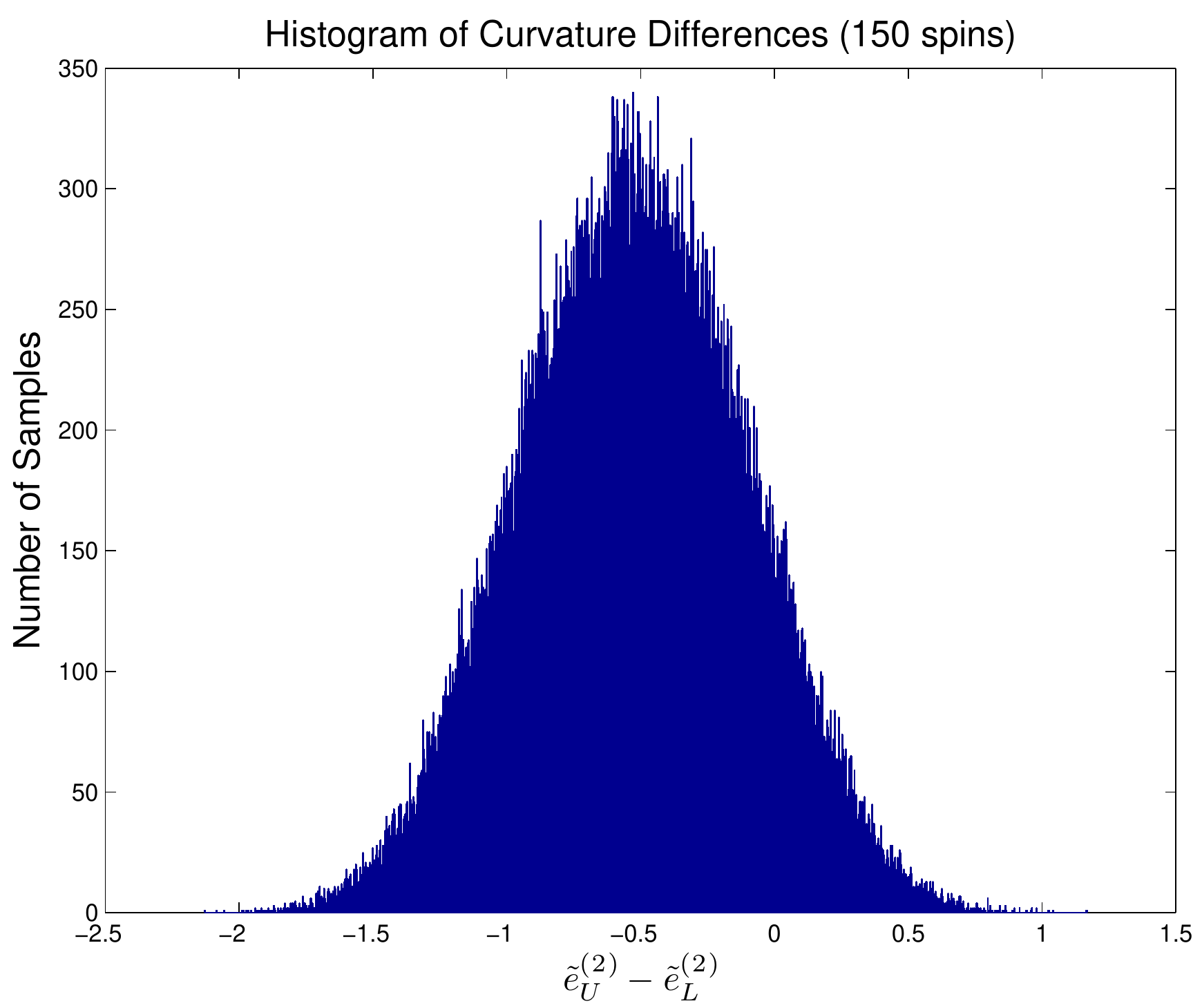}

\caption{Histogram of $\tilde{e}_{U}^{(2)}-\tilde{e}_{L}^{(2)}$ for $100000$
choices of coefficients $c_{i}$ for our 150 spin instance. Note that
a good fraction have $\tilde{e}_{U}^{(2)}-\tilde{e}_{L}^{(2)}>0$.}

\label{Flo:hist150} 
\end{figure}

\begin{figure}[p]
 \begin{raggedright} \includegraphics[scale=0.95]{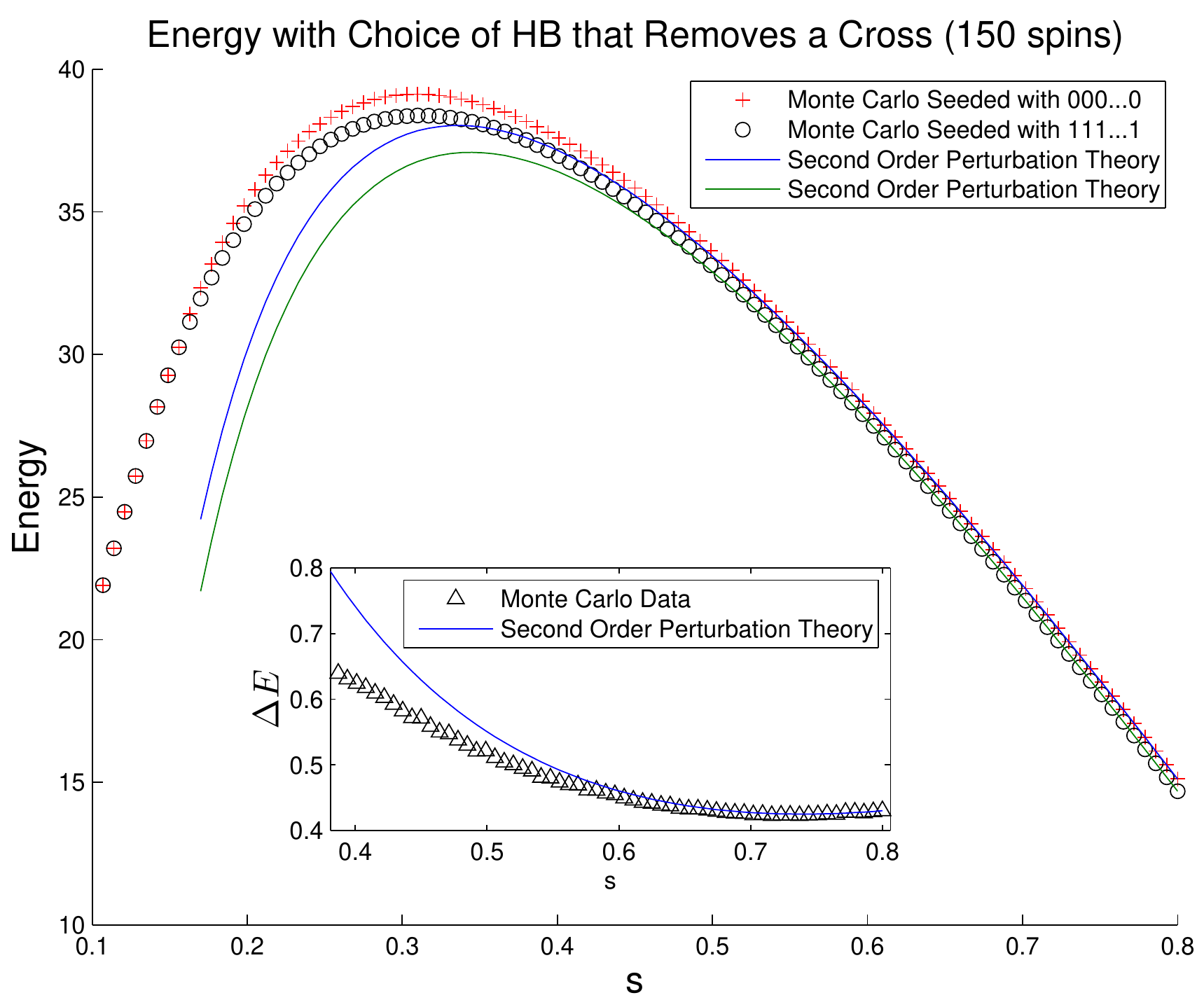} 

\end{raggedright}

\caption{A random choice of coefficients such that $\tilde{e}_{U}^{(2)}-\tilde{e}_{L}^{(2)}>\frac{1}{2}$
gives rise to an $H(s)$ where there is no longer an avoided crossing.
The circles here correspond to the ground state for all $s$ since
the cross data is always above (or equal to) the circle data for all
s. This can be seen in the inset where we have plotted the energy
difference, crosses minus circles. The crosses have a Monte Carlo
discontinuity near $s\approx0.2$, after which they correspond to
the first excited state. }

\label{Flo:150_fix} 
\end{figure}

\begin{figure}[p]
 \begin{raggedright} \includegraphics[scale=0.95]{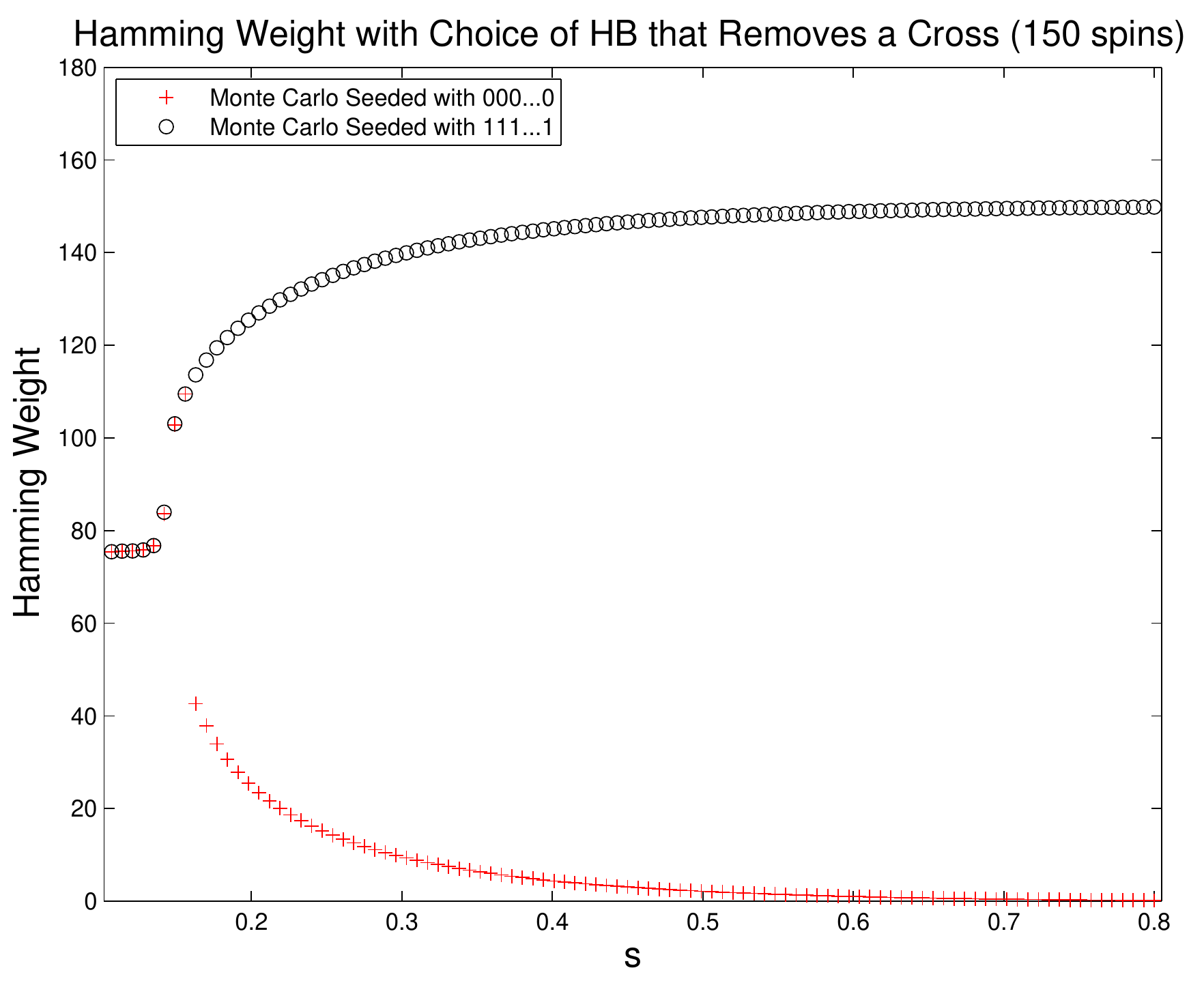} 

\end{raggedright}

\caption{Looking at figure \ref{Flo:150_fix} we see that the ground state
corresponds to the circles for all values of $s$ so we see here that
the Hamming weight of the ground state goes smoothly to its final
value as $s$ is increased. We take this as further evidence that
this choice of $\tilde{H}_{B}$ would correspond to success for the
quantum adiabatic algorithm for this instance.}

\label{Flo:150_fix_mag}
\end{figure}

The Monte Carlo data at 150 bits shows that we can make an instance
of 3SAT with a tiny gap following the procedure outlined in section
\ref{sec:Problematic-Instances}. We now use the Monte Carlo simulator
to show that a randomly chosen $\tilde{H}_{B}$ can alter the Hamiltonian
$H(s)$ so that this small gap becomes large. For the instance at
hand we first compute the difference $\tilde{e}_{U}^{(2)}-\tilde{e}_{L}^{(2)}$
for $100000$ randomly chosen sets of coefficients. The histogram
of these differences is plotted in figure \ref{Flo:hist150}. After
doing this we randomly selected (from this set) two sets of coefficients
such that $\tilde{e}_{U}^{(2)}-\tilde{e}_{L}^{(2)}>\frac{1}{2}$.
For both of these sets of coefficients we saw that the crossing at
$s\approx0.49$ was no longer present, although in one case there
appeared to be a new crossing at a much lower value of $s$. Figures
\ref{Flo:150_fix} and \ref{Flo:150_fix_mag} show the Monte Carlo
data for the choice of $\tilde{H}_{B}$ which does not appear to have
any crossing. At 150 bits we see compelling evidence that the story
outlined in sections \ref{sec:Problematic-Instances} and \ref{sec:Fixing-the-Problem}
is true.

\subsection*{Is $s^{\star}$ near $1$?}

We argued that $s^{\star}$ should go to $1$ for large enough $n$.
However our 16 bit example has $s^{\star}\approx0.42$ and the 150
bit example has $s^{\star}\approx0.49$. Recall from equation \ref{eq:s_star}
that $s^{\star}=1-\Theta(\frac{1}{n^{\nicefrac{1}{4}}}\left(\frac{m}{n}\right)^{\frac{3}{4}})$.
At 16 bits with $m=122$ we have $\frac{1}{n^{\nicefrac{1}{4}}}\left(\frac{m}{n}\right)^{\frac{3}{4}}=2.29$
and at 150 bits with $m=1783$ we have $\frac{1}{n^{\nicefrac{1}{4}}}\left(\frac{m}{n}\right)^{\frac{3}{4}}=1.83$.
Although asymptotically $m$ is of order $n\log n$, for these values
of $n$ we are not yet in the regime where $\frac{1}{n^{\nicefrac{1}{4}}}\left(\frac{m}{n}\right)^{\frac{3}{4}}\ll1$.

\begin{figure}[p]
 \begin{raggedright} \includegraphics[scale=0.95]{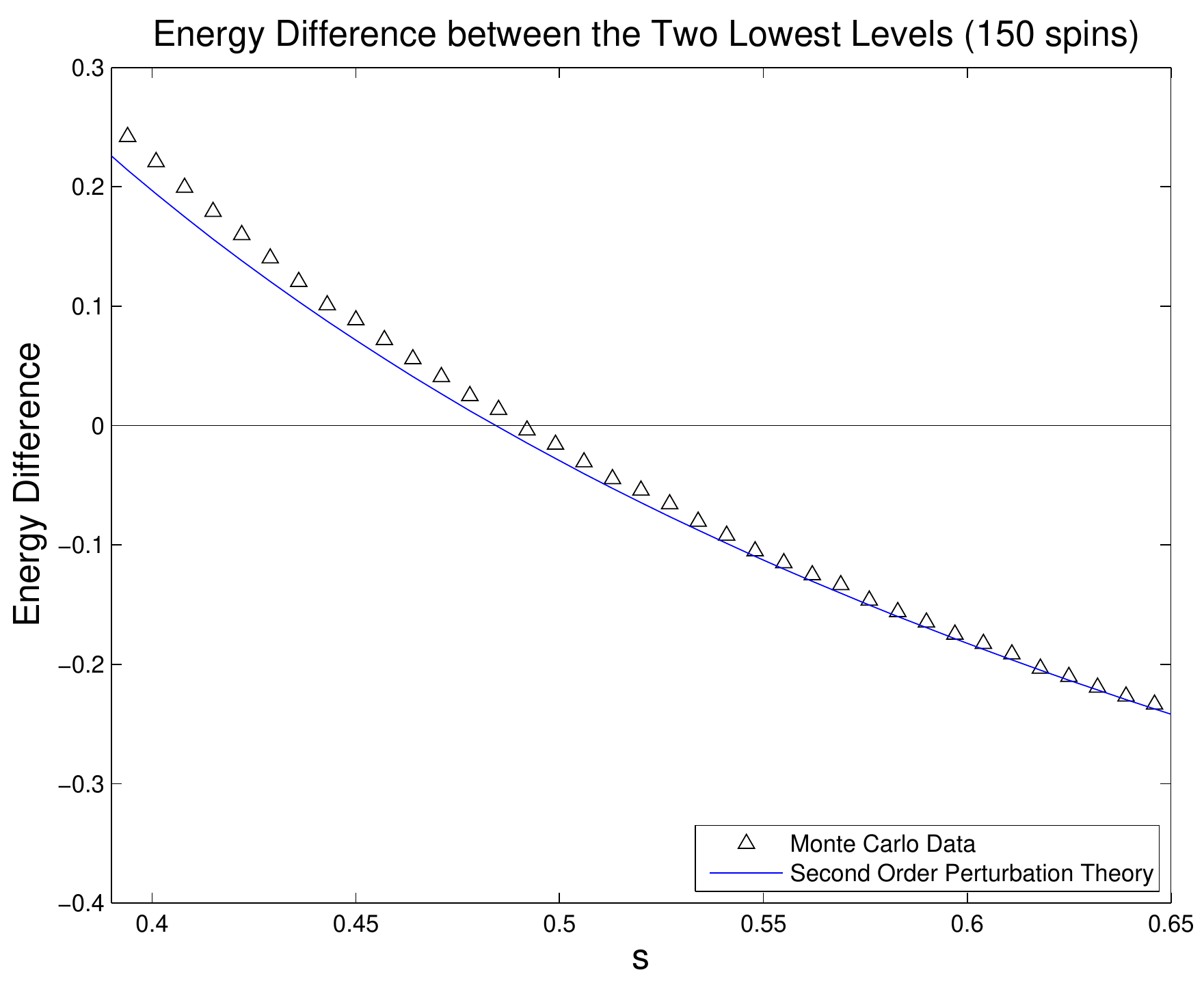} 

\end{raggedright}

\caption{The energy difference, circles minus crosses, from figure \ref{Flo:150_cross}
near the value of $s$ where the difference is $0$. Note that second
order perturbation theory does quite well in predicting where the
difference goes through zero. }

\label{Flo:150_expanded} 
\end{figure}

Even though $s^{\star}$ is not near 1 for our instance at 150 bits,
we see from figure \ref{Flo:150_expanded} that second order perturbation
theory can be used to predict the location of $s^{\star}$. This is
because the fourth order contribution to the energy difference is
quite small. So already at 150 bits we can predict the presence or
absence of an avoided crossing using second order perturbation theory.

\begin{figure}[p]
 \includegraphics[scale=0.95]{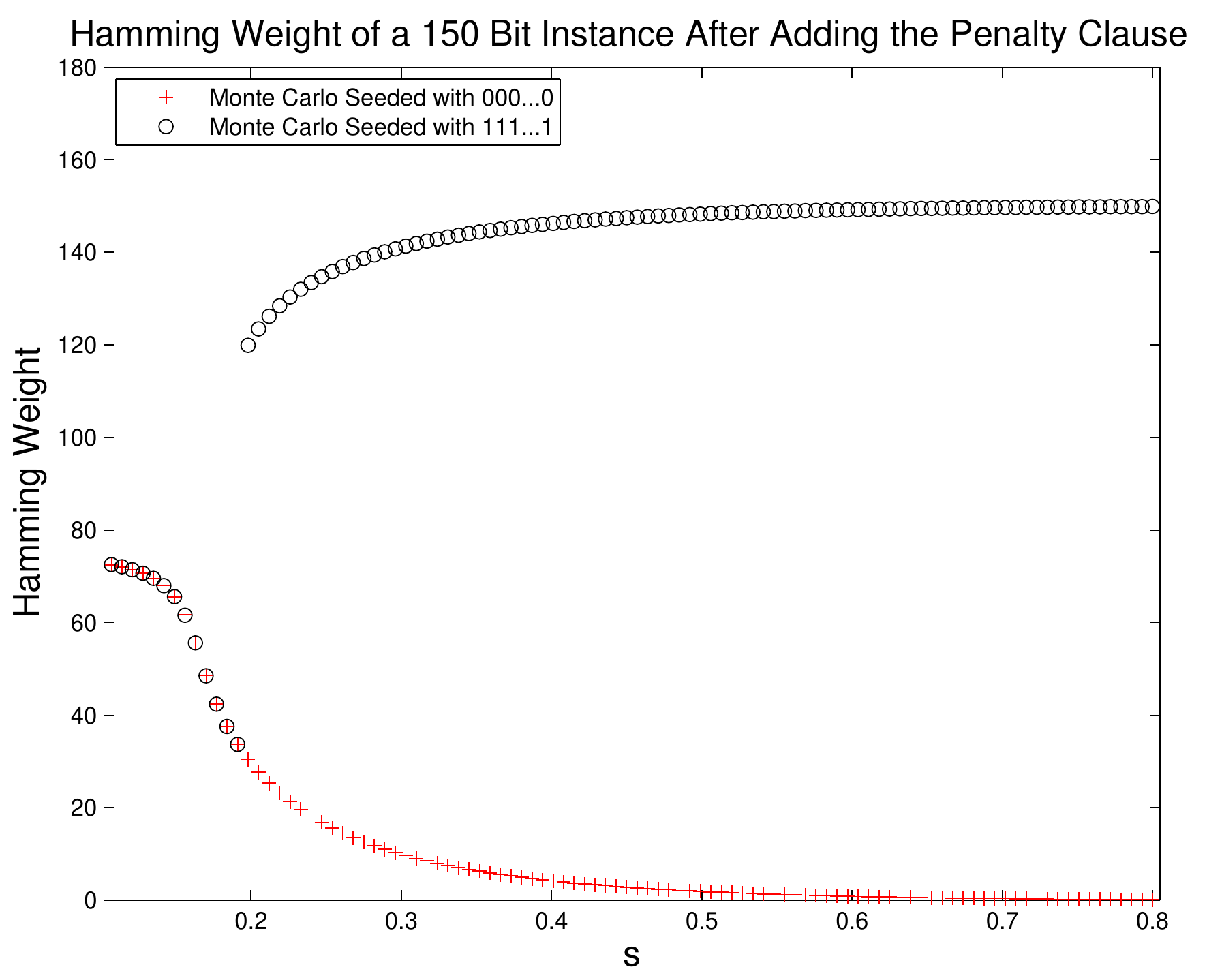}

\caption{Looking at figure \ref{Flo:150_cross} we see that the ground state
is represented by the crosses to the left of $s\approx0.49$ and is
represented by the circles after this value of $s$. Tracking the
Hamming weight of the ground state, we conclude that it changes abruptly
at $s\approx0.49$.}

\label{Flo:150_mag_cross}
\end{figure}

\section{Conclusions}
{}
We have introduced a new Quantum Monte Carlo technique
to analyze the performance of quantum adiabatic algorithms for instances of satisfiability. Using seeded configurations (in the Monte Carlo simulation) corresponding to disparate low lying states, our technique exposed the presence or absence of an exponentially small gap without ever actually computing the gap. 

We used this method to numerically investigate a set of random instances of 3SAT which were designed to expose a
weakness of the adiabatic algorithm. We confirmed that this weakness
can be overcome for our set of instances by using path change. Our numerical work and the main part of our
analysis pertains to instances of 3SAT with 2 planted satisfying assignments. 

We have also considered the scenario where an instance has $k$ satisfying assignments and then all but one are penalized with a small number of clauses. However, in the $k>2$ case we have made certain assumptions which make our analysis possible and we do not know if they apply to randomly generated instances. In our scenario the adiabatic algorithm with path change will succeed in a number of tries which is polynomially large in $k$.  Here $k$ is fixed and $n$ goes to infinity but we take the calculation in section \ref{sec:Fixing-the-Problem} as an indication that the algorithm will succeed when $k$ grows polynomially with $n$.

Throughout this paper we have assumed that our instances have a unique satisfying assignment. In this case we believe that the crucial distinction is whether there are polynomially many or exponentially many low lying mutually disparate states above the unique satisfying assignment. If there are polynomially many, we have argued that the adiabatic algorithm with path change will succeed, but if there are exponentially many we have no reason to be optimistic about the performance of the algorithm. 

When there are exponentially many satisfying assignments our analysis does not apply. This situation was considered in recent work by Knysh and Smelyanskiy \cite{knysh}.

Our results give further evidence that path change must be considered
an integral part of the quantum adiabatic algorithm. For any given
instance, the algorithm should be run with many different randomly
selected paths which end at the problem Hamiltonian. As long as the
algorithm succeeds on at least a polynomially small fraction of the
trials, it can be used to solve decision problems. 

\section{Acknowledgements}

We would like to thank Mohammad Amin, Yale Fan, Florent Krzakala,
Jeremie Roland, and Peter Young for interesting discussions. We also
thank Alan Edelman for generously offering us access to his SiCortex
computer cluster and Andy Lutomirski for fixing it. This work was
supported in part by funds provided by the U.S. Department of Energy
under cooperative research agreement DE-FG02-94ER40818, the W. M.
Keck Foundation Center for Extreme Quantum Information Theory, the
U.S Army Research Laboratory's Army Research Office through grant
number W911NF-09-1-0438, the National Science Foundation through grant
number CCF-0829421, and the Natural Sciences and Engineering Research
Council of Canada.

\bibliographystyle{plain} \nocite{*} \bibliographystyle{plain}
\bibliography{2plantbib_V2}

\appendix

\section*{Appendix: A modified Version of the Heat Bath algorithm of Krzakala
et al\label{sec:A-modified-Version}}

The authors of \cite{krzakala-2008-78} give a Quantum Monte Carlo
algorithm for spin systems in a transverse field. The algorithm we
use, which is described in this section, is a modified version of
that algorithm. The modification which we have made is described in
subsection \ref{sub:Algorithm-for-Sampling}; everything else in this
section constitutes a review of reference \cite{krzakala-2008-78}.
Like other worldline Quantum Monte Carlo techniques, the algorithm
samples the appropriate probability distribution $\rho$ (see section
\ref{sec:Quantum-Monte-Carlo}) over paths in imaginary time via Markov
Chain Monte Carlo. However this algorithm is only applicable to the
case where the Hamiltonian is of the form $H=H_{0}+V$, where $H_{0}$
is diagonal in the computational basis $|z\rangle$ and\[
V=-\sum_{i=1}^{n}c_{i}\sigma_{x}^{i}\]
 for some set of coefficients $\{c_{i}\}$ which are all positive.

\begin{figure}[H]
\begin{centering}
\includegraphics[scale=0.55]{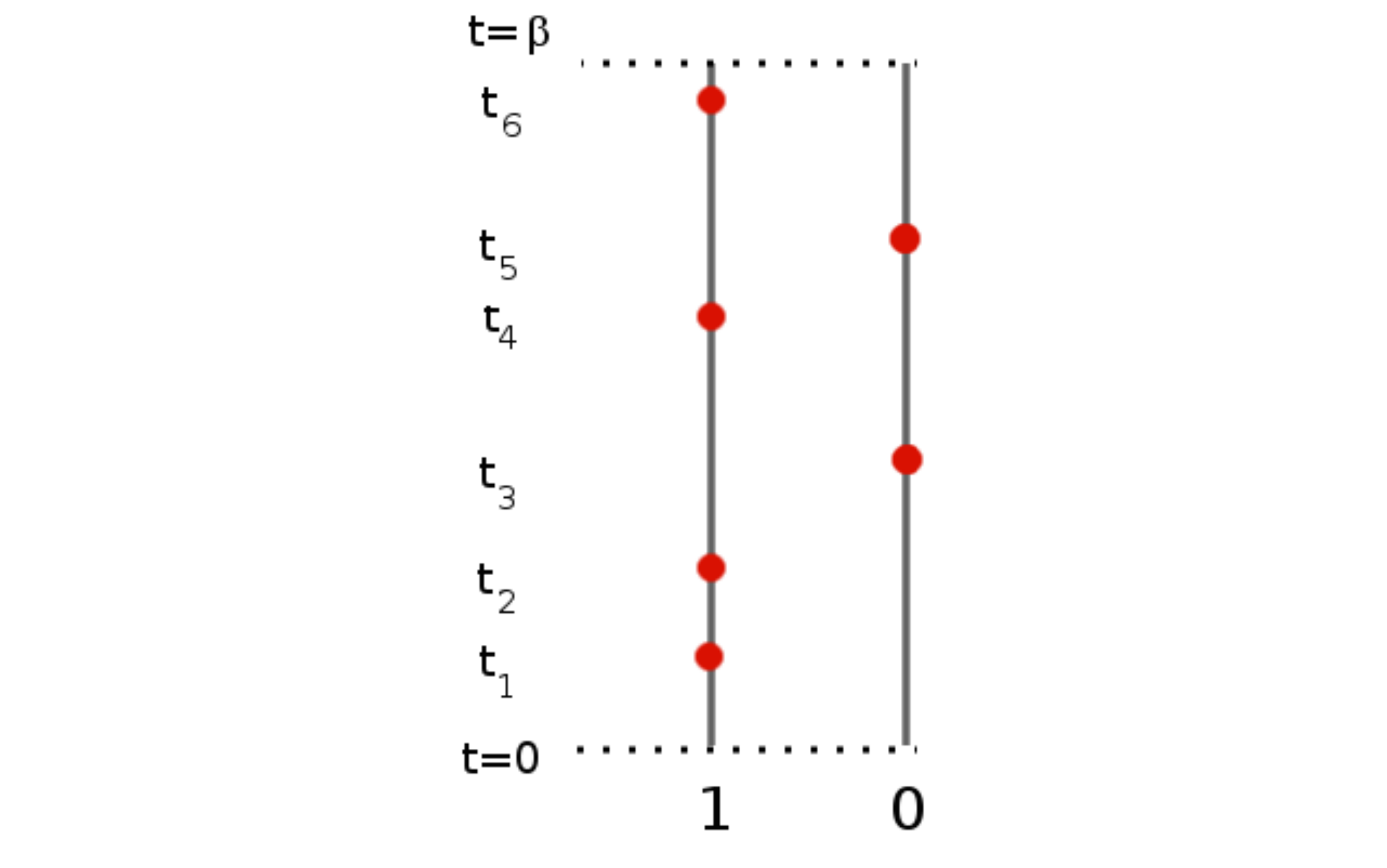}
\par\end{centering}

\caption{An example of a path P for 2 spins. The red dots indicate bit flips.
In this example the starting state $z_{1}=10$ and there are four
flips in the first spin and 2 in the second (i.e $i_{1},i_{2},i_{4},i_{6}=1$
and $i_{3},i_{5}=2$). The times of these flips are labeled $t_{1}$
through $t_{6}$.\label{fig:2spinpath}}

\end{figure}

For a Hamiltonian of this form, the distribution $\rho$ over paths
(from equation \ref{eq:rho}) is given by\[
\rho(P)=\frac{1}{Z(\beta)}\left(\Pi_{r=1}^{m}c_{i_{r}}\right)dt_{1}...dt_{m}e^{-\int_{t=0}^{\beta}\mathcal{H}_{0}(z(t))dt}\]
 where in this case a path is specified by an $n$ bit string $z_{1}$
(call this the starting state) at time $t=0$ and a sequence of flips
which occur in bits labeled $i_{1},...,i_{m}$ at times $t_{1},...,t_{m}$
(which are ordered), where each $i_{r}\in\{1,...,n\}$ and $t_{r}\in[0,\beta]$
for $r\in\{1,...,m\}$. Another way of specifying a path is to specify
the path $P_{j}$ of each spin $j\in\{1,...,n\}$. So a path $P$
of the n spin system can be written $P=(P_{1},P_{2},...,P_{n})$.
For each $j\in\{1,..,n\}$, $P_{j}$ specifies the $j$th bit of the
starting state $z_{1}$ as well as the times at which bit flips occur
in bit j. Note that we only need to consider paths which flip each
bit an even number of times, since only these paths occur with nonzero
probability. An example of a path for a system with 2 spins is given
in figure \ref{fig:2spinpath}. %

We now define in detail the Markov Chain which has limiting distribution
$\rho.$ It will be useful to define $p_{j}(P_{j}|P_{1},...,P_{j-1},P_{j+1},...,P_{n})$
(for each $j\in\{1,...,n\}$) to be the conditional probability distribution
of the path of the $j$th bit, conditioned on the remainder of the
path being fixed.

As in \cite{krzakala-2008-78}, the update rule for the Monte Carlo
algorithm consists of the following 3 steps: 
\begin{enumerate}
\item Randomly and uniformly choose a spin j (where $j\in\{1,...,n\}$). 
\item Remove all flips in the path which occur in bit $j$. Also remove
the $j$th bit from the starting state $z_{1}$. This step corresponds
to wholly removing the path $P_{j}$ of spin $j$. 
\item Draw a new path $P_{j}$ (starting value and flip times) for spin
j from the conditional distribution $p_{j}(P_{j}|P_{1},...,P_{j-1},P_{j+1},...,P_{n})$. 
\end{enumerate}
Of course, the nontrivial part of this algorithm is in specifying
a procedure which executes step (3) in the above. To do this, the
authors of reference \cite{krzakala-2008-78} note that for any index
$j\in\{1,...,n\}$, the diagonal part of the Hamiltonian can be written
as \begin{equation}
H_{0}=g_{j}+f_{j}\sigma_{z}^{j}\end{equation}
 where $g_{j}$ and $f_{j}$ are operator valued functions of all
$\{\sigma_{z}^{k}\}$ except for $\sigma_{z}^{j}$. For a given path
$P$, we define $\mathcal{F}_{j}(z(t))=\langle z(t)|f_{j}|z(t)\rangle$
for $t\in[0,\beta]$. The function $\mathcal{F}_{j}(z(t)$) is then
piecewise constant, and for a given path can be written as\[
\mathcal{F}_{j}(z(t))=\begin{cases}
\langle z_{1}|f_{j}|z_{1}\rangle, & 0\leq t<t_{1}\\
\langle z_{2}|f_{j}|z_{2}\rangle, & t_{1}\leq t<t_{2}\\
\;\quad\;\;\vdots\\
\langle z_{m}|f_{j}|z_{m}\rangle, & t_{m-1}\leq t<t_{m}\\
\langle z_{1}|f_{j}|z_{1}\rangle, & t_{m}\leq t\leq\beta\,.\end{cases}\]
 Although the above expression involves all the bit flips in the path,
the function $\mathcal{F}_{j}(z(t))$ can actually only change value
at times where bit flips occur in bits other than the $j$th bit,
since the operator $f_{j}$ does not involve $\sigma_{z}^{j}$. Let
us then write\[
\mathcal{F}_{j}(z(t))=\begin{cases}
h_{0}, & 0=\tilde{t}_{0}\leq t\leq\tilde{t}_{1}\\
h_{1}, & \tilde{t}_{1}\leq t\leq\tilde{t}_{2}\\
\:\vdots\\
h_{q}, & \tilde{t}_{q}\leq t\leq\beta=\tilde{t}_{q+1}\,.\end{cases}\]
 In this expression the times $\tilde{t}_{s}$ correspond to times
at which bit flips occur in bits other than bit $j$ (also, $h_{q}=h_{0}$).

With this notation, the procedure of reference \cite{krzakala-2008-78}
that generates a new path for bit $j$ consists of the following: 
\begin{enumerate}
\item Compute the value of $\mathcal{F}_{j}(z(t))$ as a function of imaginary
time along the path. 
\item In this step we generate boundary conditions for the path of bit $j$
at times $\tilde{t}_{0},\tilde{t}_{1},...\tilde{t}_{q}$. In other
words we choose $q+1$ values $s_{0},...,s_{q}\in\{0,1\}$ such that
at time $\tilde{t}_{r}$ the bit $j$ will be set to the value $s_{r}$
in the new path that we are generating. To do this, we sample the
values $s_{0},...,s_{q}\in\{0,1\}$ for spin j at the times $\tilde{t}_{0},...,\tilde{t}_{q}$
from their joint distribution, which is given by\[
Z(s_{0},s_{1},...,s_{q}|\{h_{0},...,h_{q}\},\{\tilde{t}_{1},...,\tilde{t}_{q}\})=\frac{\langle s_{0}|A_{q}|s_{q}\rangle\langle s_{q}|A_{q-1}|s_{q-1}\rangle...\langle s_{1}|A_{0}|s_{0}\rangle}{Tr[A_{q}A_{q-1}...A_{0}]}\]
 where \[
A_{i}=e^{-\lambda_{i}\left[h_{i}\sigma_{z}^{j}-c_{j}\sigma_{x}^{j}\right]}\]
 and $\lambda_{i}=\tilde{t}_{i+1}-\tilde{t}_{i}$ . We also define
$s_{q+1}=s_{0}$ . 
\item Having chosen boundary conditions, we now generate subpaths for bit
$j$ on each interval $[\tilde{t}_{i},\tilde{t}_{i+1}]$ of length
$\lambda_{i}$. Such a subpath is specified by a number of flips $w$
and the time offsets $\tau_{1},\tau_{2},...,\tau_{w}\in[0,\lambda_{i}]$
at which flips occur (note that the starting value of the bit is determined
by the boundary conditions). The number of flips is restricted to
be either even or odd depending on the boundary conditions that were
chosen for this interval in the previous step. In this step, the subpath
for each interval $[\tilde{t}_{i},\tilde{t}_{i+1}]$ is drawn from
the distribution \begin{equation}
g_{i}(\tau,...,\tau_{w})=\frac{1}{\langle s_{i+1}|A_{i}|s_{i}\rangle}c_{j}{}^{w}e^{-s_{i}h_{i}[(\tau_{1}-0)-(\tau_{2}-\tau_{1})+(\tau_{3}-\tau_{2})-...+(\lambda_{i}-\tau_{w})]}d\tau_{1}...d\tau_{w}\label{eq:goftau}\end{equation}

\item Put all the subpaths together to form a new path $P_{j}$ for bit
$j$ on $[0,\beta]$. 
\end{enumerate}
In section \ref{sub:Generating-Boundary-Conditions} we review the
method outlined in \cite{krzakala-2008-78} for sampling from the
distribution $Z(s_{0},s_{1},...,s_{q}|(h_{0},...,h_{q}),(\tilde{t}_{1},...,\tilde{t}_{q}))$
in step (2) of the above. In section \ref{sub:Algorithm-for-Sampling}
we outline our method for sampling from the distribution $g_{i}(\tau_{1},...,\tau_{w})$,
which differs from the method suggested in reference \cite{krzakala-2008-78}.

\subsection{Generating Boundary Conditions for A Single Spin Path\label{sub:Generating-Boundary-Conditions}}

The prescription outlined in \cite{krzakala-2008-78} for generating
a set $s_{0},s_{1},...,s_{q}$ from the distribution \[
Z(s_{0},s_{1},...,s_{q}|(h_{0},...,h_{q}),(\tilde{t}_{1},...,\tilde{t}_{q}))=\frac{\langle s_{0}|A_{q}|s_{q}\rangle\langle s_{q}|A_{q-1}|s_{q-1}\rangle...\langle s_{1}|A_{0}|s_{0}\rangle}{Tr[A_{q}A_{q-1}...A_{0}]}\]
 is as follows. First generate $s_{0}\in\{0,1\}$ according to the
distribution \[
p(s_{0})=\frac{1}{Tr[A_{q}A_{q-1}...A_{0}]}\langle s_{0}|A_{q}A_{q-1}...A_{0}|s_{0}\rangle\,.\]
 (computing these probabilities involves multiplying $q$ two by two
matrices). Then generate $s_{1}\in\{0,1\}$ according to \[
p(s_{1}|s_{0})=\frac{1}{\langle s_{0}|A_{q}A_{q-1}...A_{1}A_{0}|s_{0}\rangle}\langle s_{0}|A_{q}A_{q-1}...A_{1}|s_{1}\rangle\langle s_{1}|A_{0}|s_{0}\rangle\,.\]
 Then generate $s_{2}\in\{0,1\}$ from the distribution\[
p(s_{2}|s_{1},s_{0})=\frac{1}{\langle s_{0}|A_{q}A_{q-1}...A_{1}|s_{1}\rangle}\langle s_{0}|A_{q}A_{q-1}...A_{2}|s_{2}\rangle\langle s_{2}|A_{1}|s_{1}\rangle\]
 and so on. Note that this generates the correct distribution since\[
p(s_{0})p(s_{1}|s_{0})p(s_{2}|s_{1},s_{0})...p(s_{q}|s_{q-1},...,s_{0})=Z(s_{0},s_{1},...,s_{q}|(h_{0},...,h_{q}),(\tilde{t}_{1},...,\tilde{t}_{q}))\,.\]

\subsection{Algorithm for Sampling from the Single Spin Path Integral with Fixed
Boundary Conditions\label{sub:Algorithm-for-Sampling}}

We now present an algorithm which samples from the normalized probability
distribution over single spin paths $S(t)$ for $t\in[0,\lambda]$
($S(t)$ takes values in $\{0,1\}$). Here we only present the case
of paths with boundary conditions $S(0)=0$ and $S(\lambda)=1$. The
other three cases are completely analogous. We can parameterize such
a path with fixed boundary conditions by the times $\{\tau_{1},...,\tau_{w}\}$
at which $S(t)$ changes value. Note that the number $w$ of such
flips is odd due to our choice of boundary conditions. The distribution
over paths that we aim to sample from is given by \begin{equation}
\frac{c{}^{w}e^{-h[(\tau_{1}-0)-(\tau_{2}-\tau_{1})+(\tau_{3}-\tau_{2})-...+(\lambda-\tau_{w})]}d\tau_{1}...d\tau_{w}}{\langle1|e^{-\lambda[h\sigma_{z}-c\sigma_{x}]}|0\rangle}\,.\label{distr}\end{equation}
 It will also be useful for us to make the change of variables from
the times $(\tau_{1},...,\tau_{w})$ to the waiting times $(u_{1},...,u_{w})$
defined by

\begin{eqnarray*}
u_{1} & = & \tau_{1}\\
u_{j} & = & \tau_{j}-\tau_{j-1}\quad j\geq2\end{eqnarray*}
 Then the weight assigned to each path is\[
\frac{c{}^{w}e^{-h[u_{1}-u_{2}+u_{3}-...+\lambda-\sum_{k=1}^{w}u_{k}]}du_{1}...du_{w}}{\langle1|e^{-\lambda[h\sigma_{z}-c\sigma_{x}]}|0\rangle}\]
 which we can also write as\[
\frac{c^{w}e^{-\int_{t=0}^{\lambda}h\left(1-2S(t)\right)dt}du_{1}...du_{w}}{\langle1|e^{-\lambda[h\sigma_{z}-c\sigma_{x}]}|0\rangle}\,.\]
 The algorithm is as follows: 
\begin{enumerate}
\item Start at t=0 in state $S(0)$ defined by the boundary conditions.
Define $B_{1}=1-2S(0).$ Set i=1. 
\item Draw the waiting time $u_{i}$ until the next flip from the distribution
\[
f(u_{i})=[\sqrt{h^{2}+c^{2}}+B_{i}h]e^{-u_{i}[\sqrt{h^{2}+c^{2}}+B_{i}h]}\]
 If $\sum_{j=1}^{i}u_{j}>\lambda$ then go to step 3. Otherwise define
$B_{i+1}=-B_{i}$ and set $i\rightarrow i+1$ and repeat step 2. 
\item Take the path you have generated (which will in general be longer
than $\lambda$), and look at the segment $[0,\lambda]$. If this
path satisfies the boundary condition at $t=\lambda$ then take this
to be the generated path. Otherwise, throw away the path and repeat
from step (1). 
\end{enumerate}
We now show that this algorithm generates paths from the distribution
\eqref{distr}. Before conditioning on the boundary conditions being
satisfied, the probability of generating a sequence of waiting times
in $(u_{1},u_{1}+du_{1}),(u_{2},u_{2}+du_{2}),...(u_{w},u_{w}+du_{w})$
followed by any waiting time $u_{w+1}$ such that $u_{w+1}>\lambda-\sum_{j=1}^{w}u_{i}$
is given by\begin{eqnarray*}
 &  & f(u_{1})f(u_{2})...f(u_{w})du_{1}du_{2}...du_{w}\text{Prob}\big(u_{w+1}>\lambda-\sum_{j=1}^{w}u_{i}\big)\\
 & = & f(u_{1})f(u_{2})...f(u_{w})du_{1}du_{2}...du_{w}e^{-(\lambda-\sum_{j=1}^{w}u_{i})[\sqrt{h^{2}+c^{2}}+B_{w+1}h]}\\
 & = & \bigg[\bigg(\prod_{i=1}^{w}[\sqrt{h^{2}+c^{2}}+B_{i}h]\bigg)e^{-\sum_{i=1}^{w}u_{i}\sqrt{h^{2}+c^{2}}}e^{-\sum_{i=1}^{w}u_{i}B_{i}h}du_{1}...du_{w}e^{-(\lambda-\sum_{j=1}^{w}u_{j})[\sqrt{h^{2}+c^{2}}+SB_{w+1}h]}\bigg]\\
 & = & \begin{cases}
c{}^{w}e^{-\lambda\sqrt{c^{2}+h^{2}}}e^{-\int_{t=0}^{\lambda}\left(1-2S(t)\right)h}du_{1}...du_{w}, & \mbox{if }w\mbox{ is even}\\
c{}^{w}\left[\sqrt{1+(\frac{h}{c})^{2}}+B_{1}\left(\frac{h}{c}\right)\right]e^{-\lambda\sqrt{c^{2}+h^{2}}}e^{-\int_{t=0}^{\lambda}\left(1-2S(t)\right)h}du_{1}...du_{w}, & \mbox{if }w\mbox{ is odd }\,.\end{cases}\end{eqnarray*}
 In the last line we have used the difference of squares formula to
simplify consecutive terms: $[\sqrt{c^{2}+h^{2}}+h][\sqrt{c^{2}+h^{2}}-h]=c^{2}$.
When we condition on fixed boundary conditions (whatever they may
be), this generates the correct distribution over paths. 
\end{document}